\documentclass[%
pre,
tightenlines,
showpacs,showkeys,
a4paper,12pt
]{revtex4}
\usepackage{times}
\usepackage{amssymb,amsmath}
\usepackage{graphicx}
\usepackage{subfigure}

\renewcommand{\thesubfigure}{(\alph{subfigure})}

\makeatletter
\renewcommand{\@thesubfigure}{\thesubfigure\space}

\newcommand{\vc}[1]{\mathbf{#1}}
\newcommand{\uvc}[1]{\mathbf{\hat #1}}

\newcommand{\anch}{\mathrm{anch}}

\newcommand{\srf}{\mathrm{s}}

\newcommand{\dega}{\ensuremath{^\circ}}
\newcommand{\degc}{$^\circ$C}
\newcommand{\mum}{$\mu$m}

\makeatother

\begin{document}
\DeclareGraphicsExtensions{.eps,.jpg,.pdf}

\title{Liquid crystal anchoring transitions on aligning substrates processed by plasma beam}

\author{Oleg~V.~Yaroshchuk}
\email[Email address: ]{olegyar@iop.kiev.ua}

\author{Alexei~D.~Kiselev}
\email[Email address: ]{kiselev@iop.kiev.ua}

\author{Ruslan~M.~Kravchuk}

\affiliation{%
 Institute of Physics of National Academy of Sciences of Ukraine,
 prospekt Nauki 46,
 03028 Ky\"{\i}v, Ukraine} 

\date{\today}

\begin{abstract}
We observe a sequence of the anchoring transitions in nematic liquid crystals
(NLC) sandwiched between the hydrophobic polyimide substrates 
treated with the plasma beam. 
There is a pronounced continuous transition  from homeotropic to  low tilted
(nearly planar) alignment with the easy axis parallel to the incidence plane of the plasma beam
(the zenithal transition) that takes place 
as the exposure dose increases. 
In NLC with positive dielectric anisotropy, 
a further increase in the exposure dose results in 
in-plane reorientation of the easy axis by 90\dega\ (the azimuthal transition). 
This transition occurs through the two-fold degenerated alignment
characteristic for the second order anchoring transitions. 
In contrast to critical behavior of anchoring, 
the contact angle of NLC and water on the treated substrates 
monotonically declines with the exposure dose.
It follows that the surface concentration of hydrophobic chains
decreases continuously. 
The anchoring transitions under consideration are qualitatively
interpreted by using
a simple phenomenological model of competing easy axes
which is studied by analyzing anchoring diagrams of
the generalized polar and non-polar anchoring models.
\end{abstract}



\maketitle

\begin{figure*}[!tbh]
\centering
\resizebox{130mm}{!}{\includegraphics*{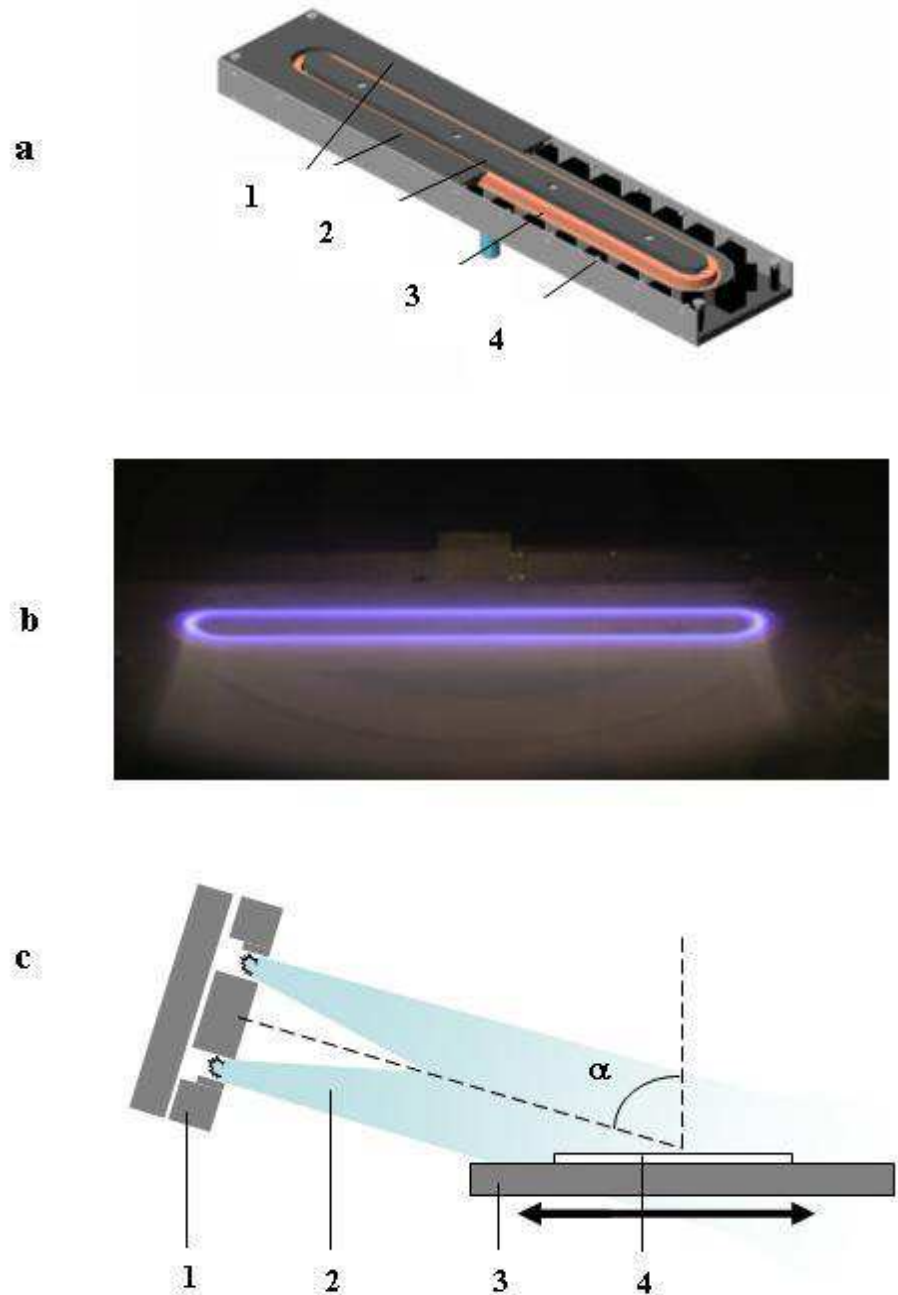}}
\caption{%
(a)~Scheme of anode layer source: (1)~inner cathode, (2)~outer
cathode, (3)~anode, (4)~permanent magnets.  
(b)~Glow discharge and beams of Ar plasma generated by anode
layer source.
(c)~Geometry of plasma beam
irradiation: (1)~anode layer source, (2)~sheet-like plasma
flux, (3)~moving platform, (4)~substrate.
}
\label{fig:anode-beam}
\end{figure*}

\begin{figure*}[!tbh]
\centering
\resizebox{150mm}{!}{\includegraphics*{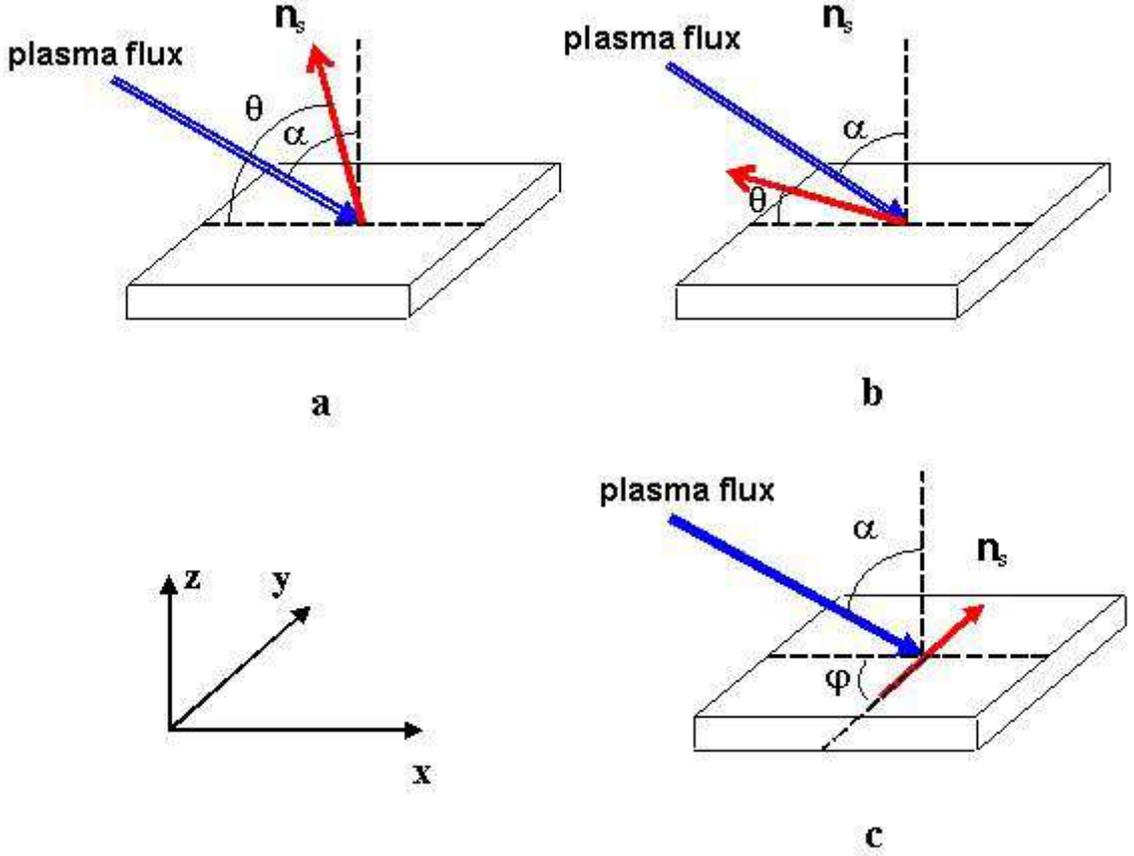}}
\caption{%
Types of NLC alignment observed in our experiments: 
(a)~low tilted (nearly homeotropic) structure 
(alignment of the 1st type with 
$75\dega\le\theta\le90\dega$ and $\phi=0\dega$); 
(b)~high tilted (nearly planar) structure 
(alignment of the 2nd type with 
$0\dega\le\theta\le30\dega$ and $\phi=0\dega$) 
which is close to planar anchoring
and (c)~planar anchoring normal to the
incidence (the $x$-$z$) plane (alignment of the 3rd type
with $\theta=0\dega$ and $\phi=90\dega$).
}
\label{fig:geometry}
\end{figure*}

\begin{figure*}[!tbh]
\centering
\subfigure[5CB]{%
\resizebox{120mm}{!}{\includegraphics*{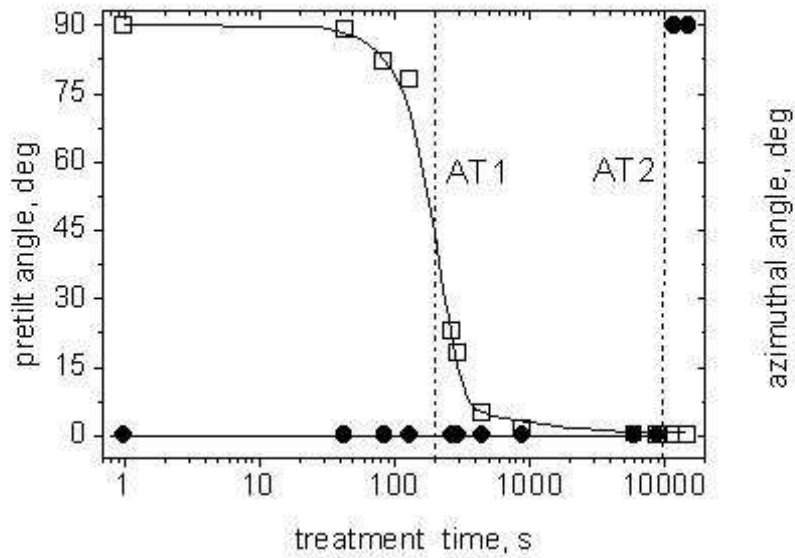}}
\label{subfig:tilt-5CB}
}
\subfigure[MJ961180]{%
\resizebox{120mm}{!}{\includegraphics*{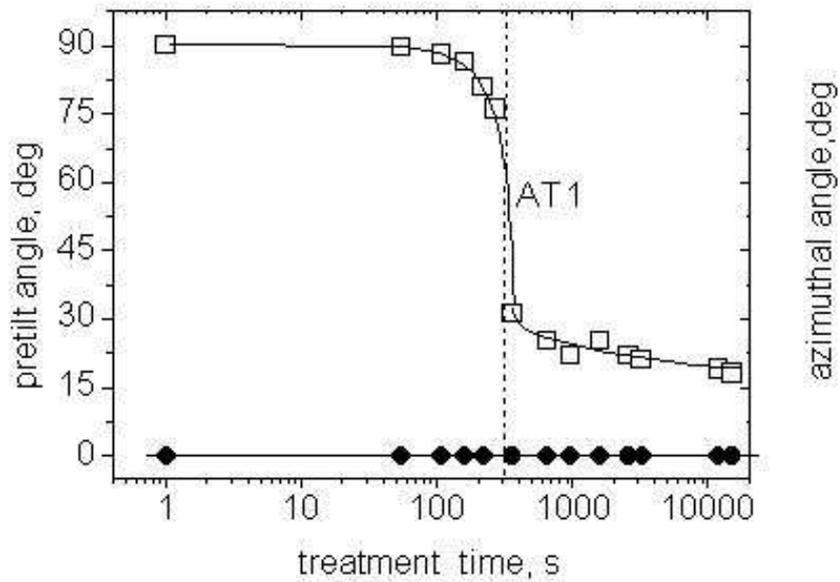}}
\label{subfig:tilt-MJ961180}
}
\caption{%
Pretilt angle, $\theta$, (open squares) and azimuthal angle, 
$\phi$, (filled circles)
measured as a function of the treatment (exposure) time 
in LC 5CB (a) and LC MJ961180 (b) at plasma-modified PI-F substrates. 
Treatment conditions are: 
$\alpha=75\dega$, j=0.4~$\mu$A~cm$^{-2}$, U=600~V.
Zenithal and azimuthal anchoring transitions are marked AT1 and AT2,
respectively.
}
\label{fig:tilt-azim}
\end{figure*}

\begin{figure*}[!tbh]
\centering
\resizebox{140mm}{!}{\includegraphics*{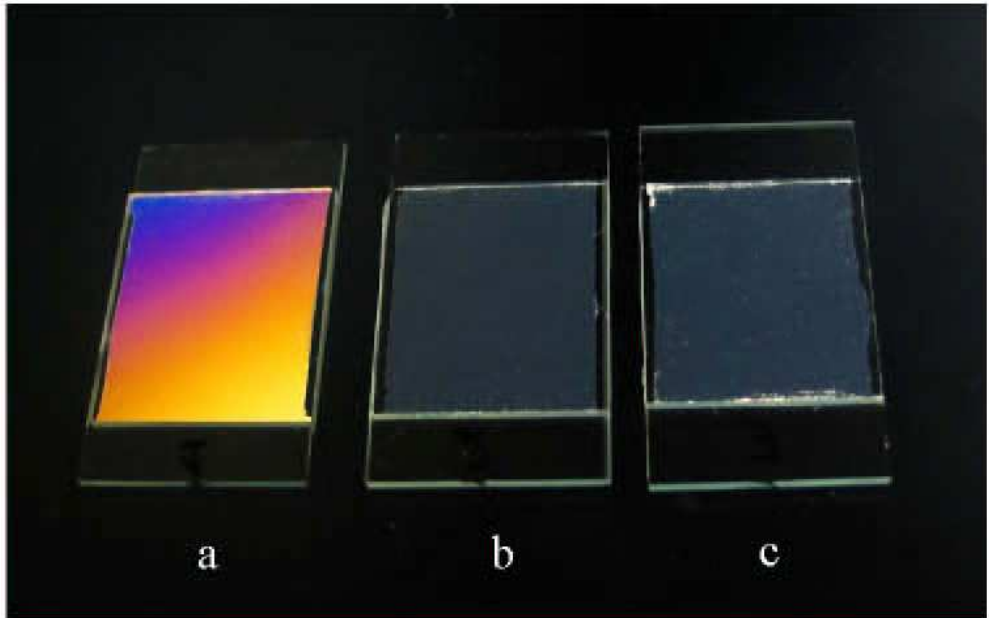}}
\caption{%
Symmetric antiparallel cells filled with LC 5CB and viewed between a pair of 
crossed polarizers. 
The PI-F substrates are treated with the plasma beam over 40 (a), 1000 (b) and
15000 (c) seconds. The cells demonstrate three
different types of alignment: (a)~high tilted, 
(b)~low tilted and (c)~planar, respectively. 
Treatment conditions are:
$\alpha=75\dega$, j=0.4 $\mu$A~cm$^{-2}$, U=600~V.
}
\label{fig:sym-cells}
\end{figure*}

\begin{figure*}[!tbh]
\centering
\resizebox{130mm}{!}{\includegraphics*{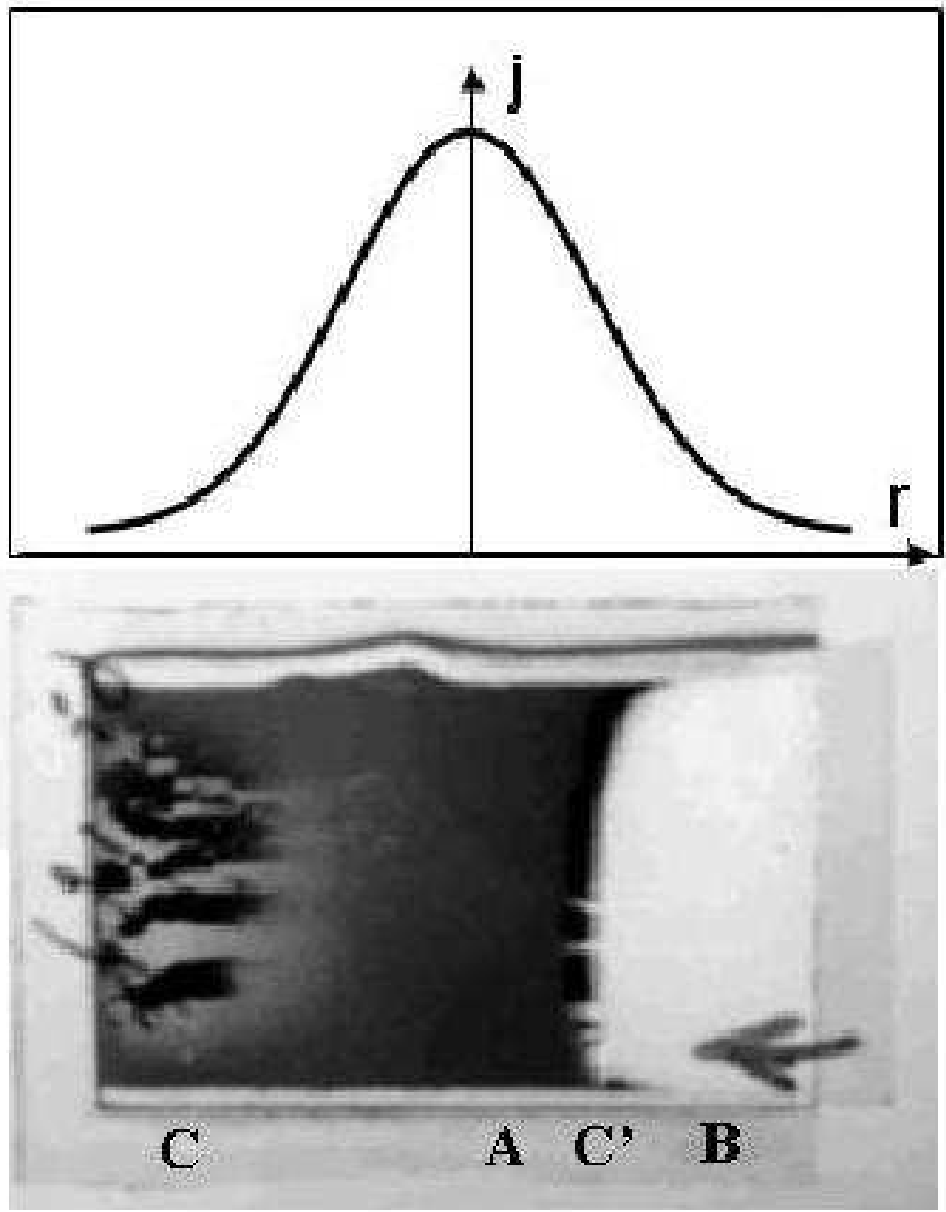}}
\caption{%
The photo of asymmetric cell viewed between parallel polarizer
and analyzer. The reference substrate is the rubbed PI film, whereas 
the tested substrate is the PI-F layer treated with the plasma beam 
in the static regime
($\alpha=75\dega$, j=7~$\mu$A~cm$^{-2}$, U=600~V, $\tau=5$~min). 
The directions of rubbing and of the plasma beam are arranged to be parallel. 
The curve of Gaussian shape, depicted above
the cell, schematically represents the distribution of the plasma beam intensity
over the tested substrate. 
In the central part of the tested substrate  irradiated at the maximum
intensity (part A),
NLC alignment corresponds to planar anchoring 
with the easy axis normal to the incidence plane (alignment of the 3rd type).
 In periphery part subjected to low irradiation doses (part B), 
the easy axis of planar anchoring lies in the plane of incidence (alignment of the 2nd type). 
These parts are separated by 
planar oriented strips (part C and part C') of transient two-fold degenerated alignment. 
}
\label{fig:asym-cells}
\end{figure*}

\begin{figure*}[!tbh]
\centering
\resizebox{150mm}{!}{\includegraphics*{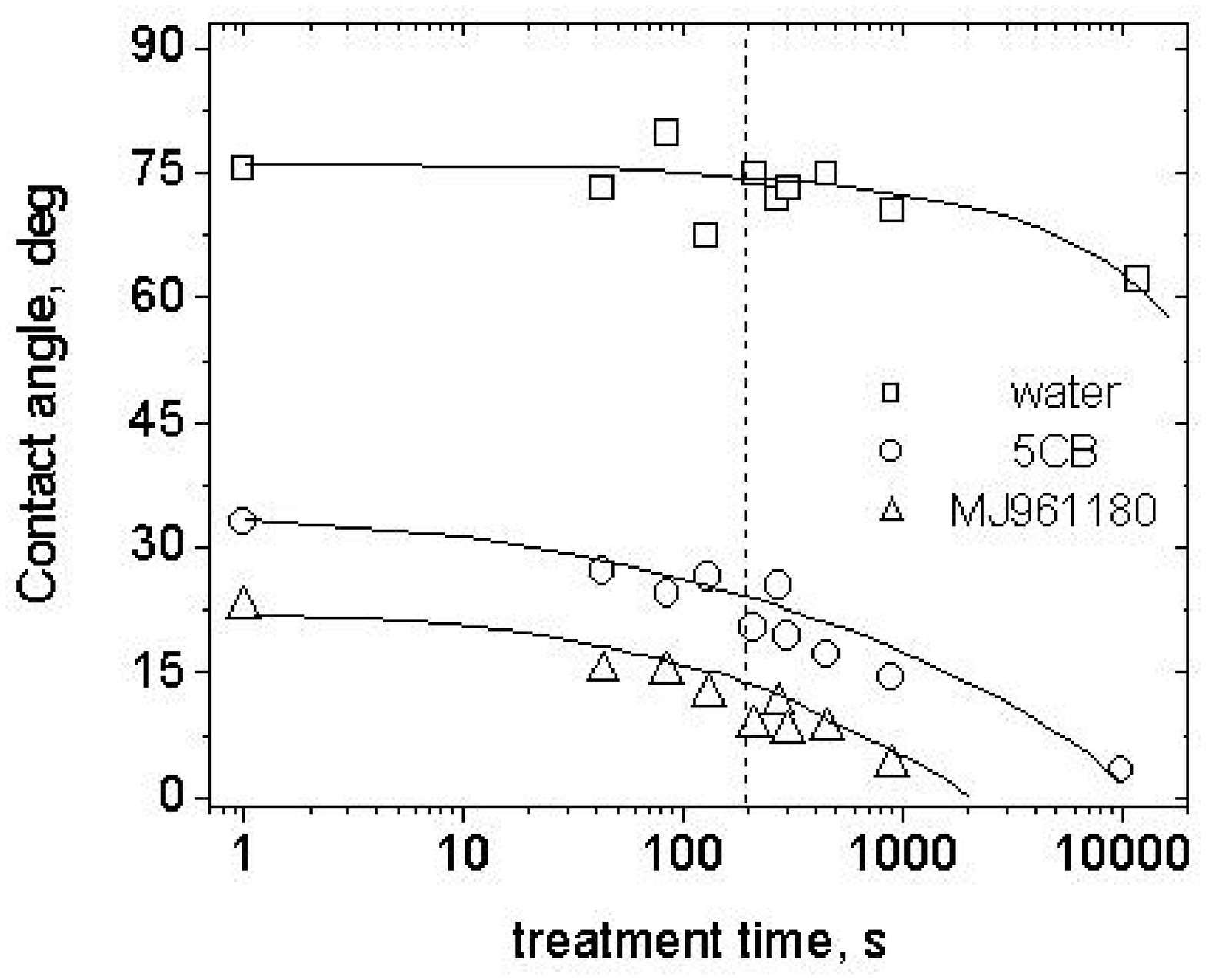}}
\caption{%
Contact angle as a function of the exposure time
for droplets of LC 5CB (circles), LC MJ961180 (triangles) and 
distilled water (squares) 
spread upon plasma treated PI-F substrates . Treatment
conditions are:
$\alpha=75\dega$, j=0.4 $\mu$A~cm$^{-2}$, U=600~V.
}
\label{fig:cnt-angle}
\end{figure*}

\section{Introduction}
\label{sec:intro}

Orientational structure of a nematic liquid crystal (NLC) placed in contact with
an anisotropic substrate is essentially determined by
the properties of the interfacial region 
where various kinds of surface induced ordering 
may exist.
Among these are smectic layering, biaxiality and 
orientational alignment
(see, e.g.,~\cite{Sluck:in:1986,Jerom:rpp:1991,Durand:1996}
for reviews). 

At the macroscopic level,  the surface induced orientation of
NLC molecules in the interfacial layer  
manifests itself as the well-known phenomenon of \textit{anchoring}.
In the case of uniaxial anisotropy, anchoring can be roughly 
described as the tendency of the nematic director 
$\vc{n}$, to align along the direction of preferential anchoring 
orientation at the surface.
The direction of surface induced alignment is specified by
a unit vector $\vc{n}_{\srf}$ referred to as the \textit{easy axis}.

Anchoring is governed by
the so-called \textit{anchoring energy}, $W_{\anch}$, 
which is the orientationally dependent (anisotropic) part of the surface
tension. In particular, easy axes can be found by 
minimizing the anchoring potential
and, thus, crucially depend on the shape of $W_{\anch}$. 

When the anchoring energy changes,
the easy axes may vary in both direction and number.
Such variations of the anchoring conditions
result in reorientation of the NLC director
known as the \textit{anchoring (surface) transition}.

Since the anchoring potential is sensitive to the thermodynamic parameters,
the anchoring transitions, similar to the phase transitions,
can be driven by temperature, chemical potential and strain.
They can also be first and second order depending on 
whether the anchoring induced reorientation is 
discontinuous (jump-like) or continuous at the critical point.
For planar interfaces, the transitions that occur through
out-of-plane, in-plane and mixed director reorientation may be
classified as the \textit{zenithal}, \textit{azimuthal} and \textit{mixed} anchoring
transitions, respectively.

For example, a discontinuous zenithal transition from planar to homeotropic
orientation was found to occur
at a flat glass or quartz substrate on cooling toward the
smectic-A transition temperature~\cite{Kanel:pra:1981}
and on the surface of a self-assembled
monolayer, which is made sufficiently hydrophobic~\cite{Alkhair:pre:1999}.
By contrast, the temperature-driven zenithal transitions observed 
at the free NLC surface~\cite{Chiarelli:jpf:1983,Chiarelli:pla:1984,Sonin:pre:1995}
and at the rubbed polyimide aligning layers~\cite{Shio:pre:2003}
turned out to be continuous.

Transitions between different anchorings can be generated
by changing either the molecular characteristics of NLC materials 
or the parameters determining the structure of substrates.
The series of the azimuthal anchoring transitions 
on the cleaved surfaces of some crystals such as gypsum and mica
studied in relation to the composition of the
atmosphere in water and alcohol vapors above the nematic 
film~\cite{Jerome:pra:1989,Jerome:pra:1990,Jerome:prl:1990,Jerome:phtr:1991}
represent such transitions.

Of particular interest are the transitions governed by  the parameters 
that characterize the method employed to treat the surface for 
fabrication of aligning films.
A variety of photo-induced orientational surface transitions that
have been observed 
in~\cite{Gibbon:nat:1991,Li:lc:1995,Andrien:jetp:1997,Andrienko:mclc:1998,Stri:2000}, 
are related to the photoalignment technique, 
in which an aligning layer is irradiated with actinic light 
(see~\cite{Kelly:jpd:2000,Chigr:rewiev:2003}
for recent reviews).

Another approach suggested in~\cite{Jannin:apl:1972,Urbach:apl:1974}
is to align liquid crystals by obliquely evaporated thin films of silicon oxide SiO$_x$.
Anchoring of nematics at the obliquely evaporated SiO$_x$
was studied as a function of the evaporation angle~\cite{Jerome:eurpl:1988} and the film
thickness~\cite{Monkade:eurpl:1988}.
It was found that an increase in either of these parameters may initiate
the sequence of  mixed and zenithal continuous surface transitions between three
different anchorings: planar monostable, tilted bistable and tilted monostable.


In this paper we deal with anchoring transitions on the
substrates treated with ion/plasma beams. 
Recently this kind of treatment has aroused considerable interest 
because it offers the greatest promise to replace
the traditional rubbing technique in the new 
generation of liquid crystal displays 
(LCD)~\cite{Chau:nat:2001,Yarosh:lc:2004,Yarosh:sid:2005}.
This processing avoids direct mechanical
contact with aligning substrates thus minimizing the 
surface deterioration. 
It also provides highly uniform alignment on
microscopic and macroscopic scale with a widely controlled pretilt
angle and anchoring energy. 

We apply this method to treat the films
of hydrophobic polyimide and investigate the anchoring transitions 
at the plasma-modified substrates as a function of the irradiation dose. 

The layout of the paper is as follows.
Experimental procedure is described  in Sec.~\ref{sec:experim}.
We present our results in Sec.~\ref{sec:results}
and, in Sec.~\ref{sec:discussion}, discuss how they can be interpreted
theoretically using the phenomenological model of two competing easy
axes with the anchoring potential taken in the Sen-Sullivan form~\cite{Sen:1987}.
Concluding remarks are given in Sec.~\ref{sec:concl}.

\section{Experimental}
\label{sec:experim}

\subsection{Setup for plasma beam exposure }
\label{subsec:setup}

The irradiation set up was based on anode layer source (ALS) from the
Hall family of sources working in the beam mode~\cite{Zhurin:psst:1999}. 
The general
construction of this source is presented in Fig.~\ref{fig:anode-beam}(a). A glow
discharge is initiated in the crossed electric and magnetic fields
within the discharge channel formed by inner and outer cathodes and
anode. Because of high anode potential, the ions of plasma are pushed
out of discharge area. They involve electrons so that the beam of
accelerated plasma is formed. In contrast to the Kaufman source widely
used for the ion beam alignment processing~\cite{Chau:nat:2001,Hwang:jjap:2002}, 
ALS does not
contain any grids and hot elements (such as filaments and other secondary
electron sources). The structure is thus simple and allows one to
substantially increase reliability.

We used ALS with a race track shaped glow discharge so that the
source generates two "sheets" of accelerated plasma
(Fig.~\ref{fig:anode-beam}(b)). 
As we have shown previously, this construction suits very well for the alignment
treatment of large-area substrates: in case the substrate is moved
across the plasma ``sheet'', the only limiting factor for the width of this substrate is 
the width of the ``sheet''. 
Since ALS can be easily scaled up, this
process can be employed in manufacturing of modern LCD fabs 
($1870\times 2200$~mm$^2$ in the 7th generation fabs).

The feed gas was argon. The working pressure, $P$, in our experiments was
$1.4\times 10^{-4}$~Torr
that corresponded to the current density, $j$, within 
the beam 0.4~$\mu$A~cm~$^{-2}$. 
The low current was used to vary gradually the exposure dose
given by the product of the current density and 
the exposure (treatment) time, $\tau_{\text{exp}}$.
The anode potential $U$ determining the maximum energy of 
plasma Ar$^+$ ions was 600~V.

The geometry of exposure is shown in Fig.~\ref{fig:anode-beam}(c). 
The substrates were irradiated obliquely and 
the incidence angle of plasma beam, $\alpha$, was about
75\dega. The substrate's holder was mounted on the PC controlled
translator in a vacuum chamber under the discharge channel. The
substrates were treated in dynamic and static regime as well. Due to
translations, different parts of the sample were passing through the
plasma beam many times undergoing alignment treatment repeatedly (the
cycling regime of treatment) so that alignment uniformity was
substantially improved. The approximate distance between the plasma
outlet and the substrate of size of $20\times 30$~mm$^2$ was 8~cm.

\subsection{Samples and their characterization}
\label{subsec:sampl}

We used the fluorinated polyimide (PI-F) containing hydrophobic side
chains as a polymer material. The polymer was dissolved in an
appropriate solvent and spin coated on the glass plates over indium
tin oxide (ITO) electrodes. The substrates were then baked at 180\degc\
over 1.5~h to remove the solvent.

Two types of NLC cells were prepared:  
(1)~identical substrates with the
plasma treated PI-F films were assembled to form symmetric NLC cells
with antiparallel director orientation; 
(2)~the tested substrate with the plasma-modified PI-F layer and 
the reference substrate with the 
rubbed polyimide (PI) layer (9203 from JSR) were
arranged so as to form asymmetric NLC cells where  
the rubbing direction was antiparallel to the direction of plasma irradiation.
In both cases the cell thickness was kept at 20~\mum. 

The symmetric cells were used to measure the pretilt angle of NLC
by the crystal rotation method, whereas the asymmetric cells served to
determine in-plane direction of the easy axis. 
The NLCs 5CB and MJ961180
(both from Merk) in an isotropic phase were injected into the cells by
capillary action. 
The LC 5CB with positive dielectric anisotropy 
$\Delta\epsilon$ is a well characterized nematic cyanobiphenyl used as a component of
industrial TN LC mixtures. The mixture LC MJ961180 with 
$\Delta\epsilon<0$ is developed for VA LCD. 
The quality of sample alignment was judged by 
observation in polarizing microscope and with a naked eye by placing
a sample between crossed polarizers.

\section{Results}
\label{sec:results}

Referring to Fig.~\ref{fig:geometry},
orientation of the easy axis
induced with the plasma beam processing
is specified by 
the pretilt and azimuthal angles, $\theta$ and $\phi$.
Fig.~\ref{fig:tilt-azim}(a) and Fig.~\ref{fig:tilt-azim}(b)
show these angles measured as a
function of the exposure time in 
the cells filled with LC 5CB and
LC mixture MJ961180, respectively.

For the 5CB cells, the curves  indicate a pronounced 
homeotropic-to-oblique 
anchoring transition that occurs at low irradiation dose. In
this case, the easy axis initially directed along  the normal to
the substrate (the $z$ axis) inclines continuously in the incidence plane of plasma beam 
(the $x$-$z$ plane) towards the plasma beam direction (Fig.~\ref{fig:geometry}). 
When increasing the exposure time $\tau_{\text{exp}}$,
the pretilt angle first decreases gradually from 90\dega\ to
75\dega. The angle declines steeply to $\theta\approx 25$\dega\
at the critical point.  
Then it decays to the value about 2\dega\ which
weakly changes with the exposure time. 

>From dependence of the azimuthal angle on the irradiation time
plotted in Fig.~\ref{fig:tilt-azim}(a) it can be inferred
that the above zenithal transition is followed by
the azimuthal transition which takes place in the region of
long-time treatment.
In this case the result of drastic in-plane reorientation
is that the easy axis lying initially in the plane of incidence
is rotated through 90 degrees.
Thus, we have the transition between two planar anchorings:
$\vc{n}_{\srf}=\uvc{x}$ and $\vc{n}_{\srf}=\uvc{y}$
(see Fig.~\ref{fig:geometry}). 

So, the results for LC 5CB
representing nematic materials of positive
dielectric anisotropy clearly indicate two
anchoring transitions driven by the irradiation dose:
zenithal and azimuthal.
The sequence of transitions involves
three different anchorings
that can be described as three types of LC alignment: 
(1)~high tilted structure (nearly homeotropic) with
zero azimuthal angle (alignment of the 1st type) observed 
in the region of low irradiation doses before the zenithal 
transition; 
(2)~low tilted structure (nearly planar) with
zero azimuthal angle (alignment of the 2nd type) 
observed between the  anchoring transitions; 
(3) planar anchoring with the easy axis normal
to the incidence plane (alignment of the 3rd type) 
detected above the critical dose of the azimuthal transition. 
Fig.~\ref{fig:sym-cells} shows that alignment of the above listed
orientational structures is of excellent quality.

The curves presented in Fig.~\ref{fig:tilt-azim}(b) 
were measured in the cells filled with LC mixture MJ961180, 
which is a nematic material with negative dielectric anisotropy,
$\Delta\epsilon<0$.
It can be seen that, as far as the zenithal transition is concerned,
the results for this mixture are quite similar to those
obtained for 5CB cells. 
Quantitatively, as opposed to LC 5CB,
the pretilt angle above the critical point remains approximately
constant varying in the range between 30\dega\  and 15\dega\ .
The most important difference is that
the azimuthal anchoring transition 
with in-plane reorientation towards the normal of the incidence plane 
turned out to be suppressed.

The experimental data presented in Fig.~\ref{fig:tilt-azim}(a)
are insufficient to judge the character of the azimuthal transition 
unambiguously.
In order to clarify behavior of anchoring near 
the critical point AT2,
we used the substrates treated in the static regime of
irradiation. Since the beam profile in the transverse direction 
has the Gaussian shape, the exposure dose appears to 
be continuously distributed over the substrate area. 

In Fig.~\ref{fig:asym-cells}, 
the sample as viewed between parallel polarizers
is presented for  
a typical asymmetric 5CB cell with the PI-F substrate processed 
in the static regime. 
It can be concluded that, in the central part (part A) of the cell
exposed to the highest dose with the maximum intensity,
anchoring is planar with the director normal to the incidence plane
(alignment of the 3rd type).
By contrast, periphery part (part B) of the cell is characterized by planar alignment of 
the 2nd type (the easy axis is parallel to the plane of incidence).
 
There are two transient strip-like regions between the parts of low and high
irradiation doses shown in Fig.~\ref{fig:asym-cells} as parts C and C'.
These regions are divided into narrow domains.
Owing to the mirror symmetry, domains
oriented symmetrically with respect to the plane of incidence
are equiprobable provided that the irradiation dose is fixed.
 
The ``strips'' are found to differ in width.
The reason is that, for oblique irradiation, 
plasma fluxes impinging on the
upper and lower parts of the substrate
are necessarily
different in intensity magnitude and distribution shape. 

The upper strip is narrow and, as a consequence, 
the intensity is tightly distributed over the domain
with the irradiation dose varying within narrow limits.  
So, producing a substrate aligned as this strip
in the dynamic regime of irradiation 
can be  rather difficult as it requires 
using a fine tuning procedure for irradiation doses.


\section{Discussion}
\label{sec:discussion}

We can now take a closer look at the properties of 
the anchoring transitions described in the previous section.
Our first remarks concern the character of the transitions.

In our experiments, 
the   irradiation dose driven 
zenithal transition was found to 
take place in the plane of incidence for either sign of 
the NLC dielectric anisotropy.
When increasing the dose, it occurs through 
homeotropic to (nearly) planar reorientation of the easy axis
and manifests itself as a steep decline of 
the pretilt angle in the immediate vicinity of the critical dose
(see the curves in Fig.~\ref{fig:tilt-azim}).
Since reorientation does not show any discontinuities, 
it may be concluded that the transition is second order.
Transitions of this type were previously obtained
at films modified with
actinic light~\cite{Andrien:jetp:1997,Li:lc:1995}
and cold plasma~\cite{Hubert:epjb:1999,Fonseca:jap:2003,Jang:jjap:2006}.

Anchoring is monostable planar 
in 5CB cells treated for so long that 
irradiation doses are well beyond the critical point of 
the homeotropic-to-planar transition.
As it can be seen from Fig.~\ref{subfig:tilt-5CB},
the curve for the azimuthal angle suggests that
the zenithal transition, AT1, is followed by the azimuthal one, AT2.

This is the transition between two planar anchorings
in which the easy axis is either parallel or normal
to the incidence plane (the $x$-$z$ plane).
It is characterized by in-plane reorientation
of the director which is rotated abruptly by 90 degrees
near the critical point. 

But the image of the asymmetric cell
with one of the substrates treated in the static regime
(see Fig.~\ref{fig:asym-cells}) clearly shows
the presence of planar oriented domains 
where the director is tilted with respect to the plane of 
incidence. 
Such strips of transient alignment are typical of
second order transitions where fluctuations create domains having
close orientations~\cite{Jerom:rpp:1991,Jerome:phtr:1991}. 
The transitions of similar character 
were generated at obliquely evaporated SiO$_x$ 
films~\cite{Jerome:eurpl:1988,Monkade:eurpl:1988} 
and at photoaligned layers~\cite{Andrienko:mclc:1998}.

The changes of anchoring directions are caused by surface modification of the
aligning films induced by plasma beam treatment. 
By analogy with other plasma processes~\cite{Hubert:epjb:1999,Jang:jjap:2006}, 
plasma beam may destroy side hydrophobic chains
and increase the free energy of the aligning layer.
 
Gradual reduction of hydrophobic chains on the polymer surface was directly
detected by XPS (x-ray photoelectron spectroscopy) 
method in~\cite{Hubert:epjb:1999}.
We carried out the contact angle measurements 
that, according to~\cite{Akiyama:jjap:2001},  
can be used to obtain indirect experimental evidence
that the hydrophobic chains concentration diminishes
with the exposure dose. 

Fig.~\ref{fig:cnt-angle} presents the contact angle as a function of 
the exposure time measured at room temperature for 
three kinds of material: LC 5CB, LC MJ961180 and distilled water. 
It is clear that, for all compounds, 
the contact angle gradually declines with
the exposure dose. It means that surface hydrophobicity 
monotonically decreases, whereas the surface free energy increases. 
In contrast to NLC alignment, 
the contact angles do not reveal any signs of critical behavior 
at the exposure doses corresponding to AT1 and AT2.

It is reasonable to assume that the critical concentration of hydrophobic
chains should be reached to 
trigger the zenithal anchoring transition. 
This concentration is associated with the critical value of the surface free energy. 

\begin{figure*}[!tbh]
\centering
\resizebox{150mm}{!}{\includegraphics*{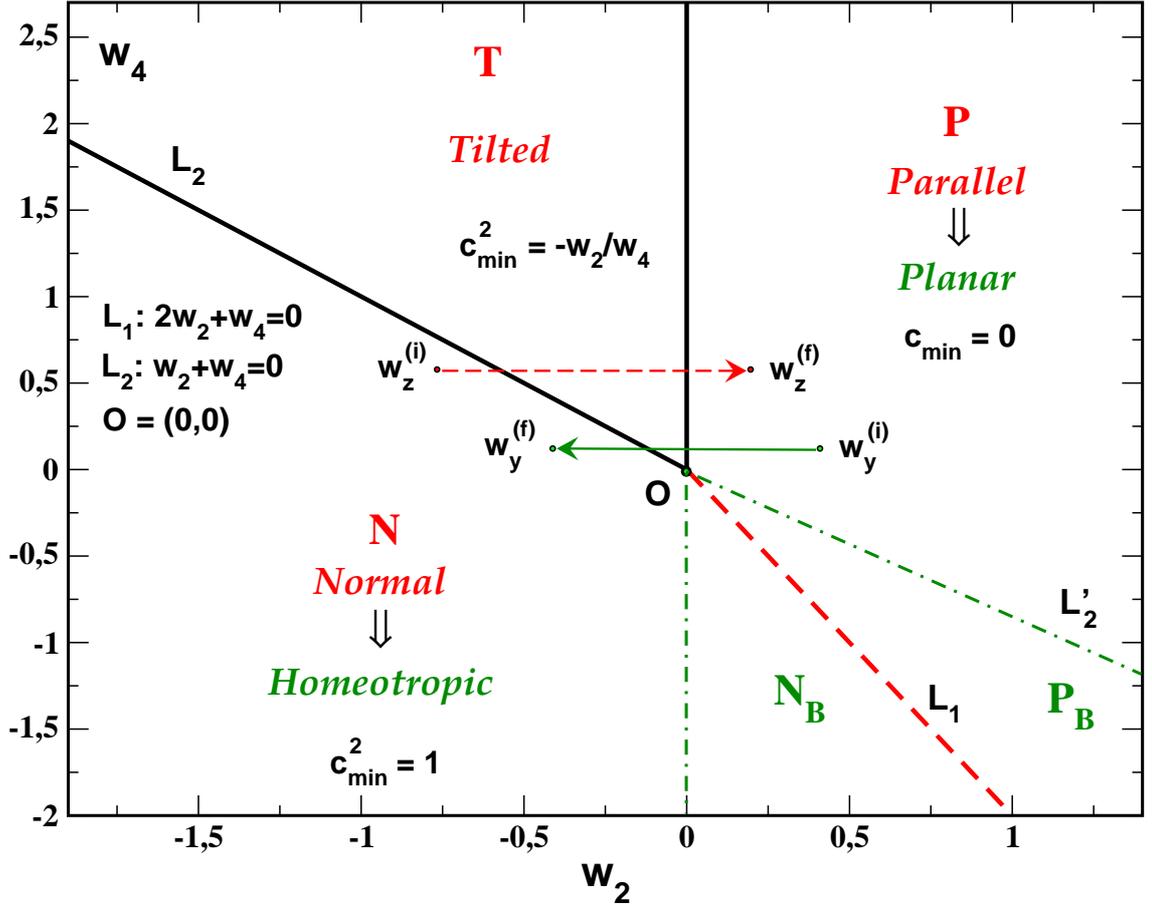}}
\caption{%
Anchoring phase diagram 
in the $w_2$--$w_4$ plane 
for the non-polar potential~\eqref{eq:gen-Parson} with $w_1=0$.
Solid and dashed lines represent continuous and discontinuous
transitions, respectively.
The subscript B identifies the regions
where the homeotropic  and planar structures are separated 
by the energy barrier.
}
\label{fig:nonpolar}
\end{figure*}

\begin{figure*}[!tbh]
\centering
\resizebox{140mm}{!}{\includegraphics*{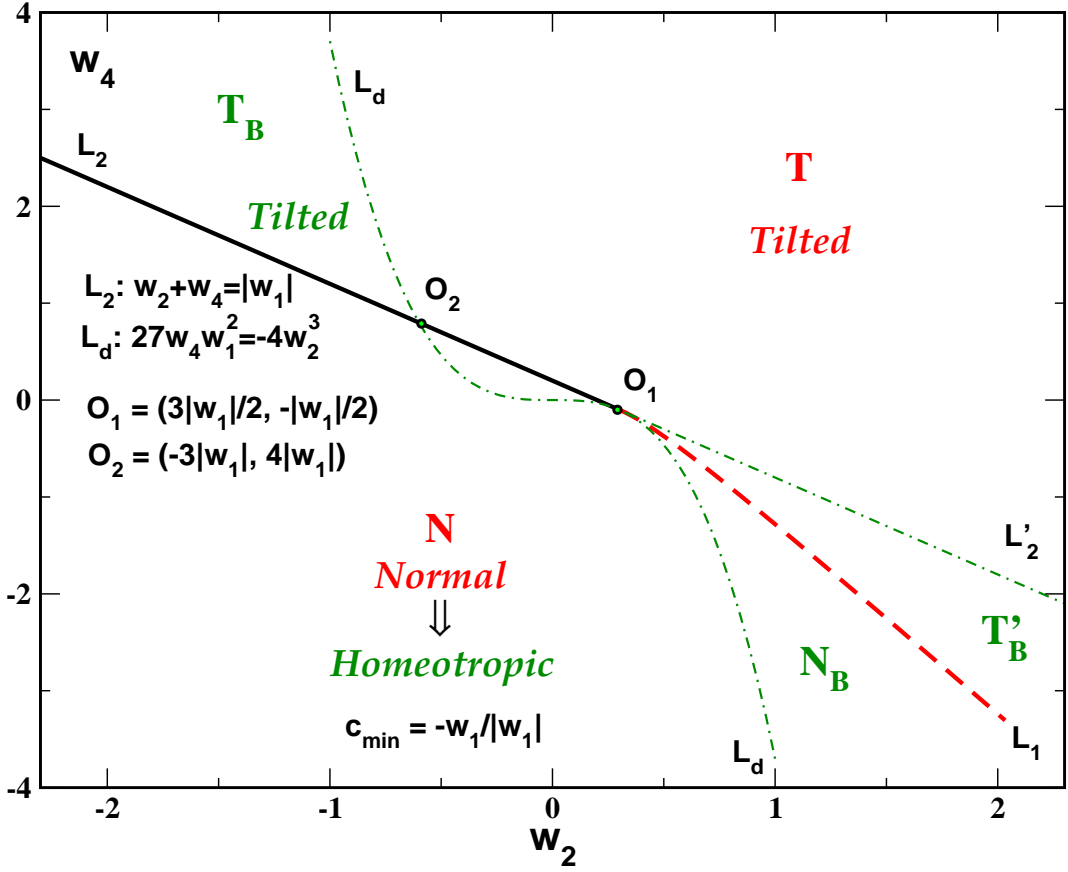}}
\caption{%
Anchoring phase diagram 
in the $w_2$--$w_4$ plane 
for the potential~\eqref{eq:gen-Parson} 
in the presence of the polar term
proportional to $w_1\ne 0$.
There is a metastable state separated from
the equilibrium structure by the energy barrier
in the regions labeled by subscript B.
The special case where 
the metastable state corresponds to 
the homeotropic anchoring is marked by prime.
}
\label{fig:polar}
\end{figure*}


\begin{figure*}[!tbh]
\centering
\resizebox{140mm}{!}{\includegraphics*{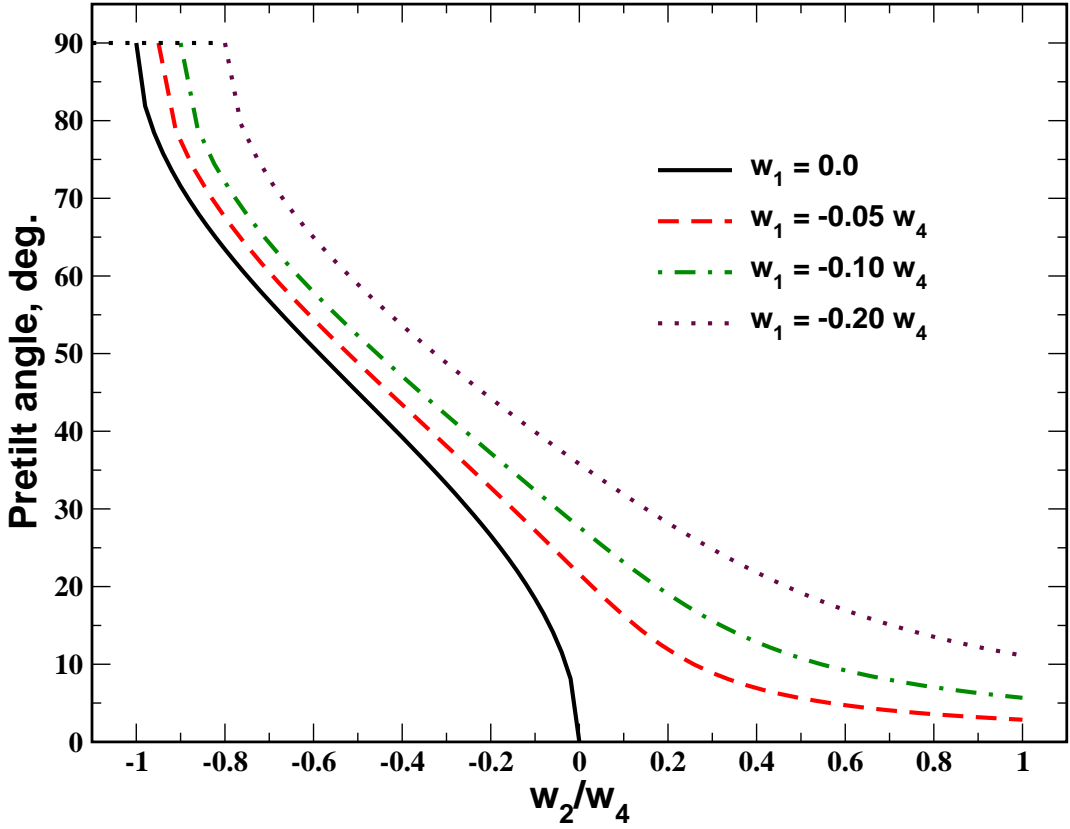}}
\caption{%
Pretilt angle versus the dimensionless anchoring parameter 
$w_2/w_4$ at different values of the polar coefficient $w_1$
for $w_4>0$.  
}
\label{fig:pretilt-theor}
\end{figure*}

\begin{figure*}[!tbh]
\centering
\resizebox{140mm}{!}{\includegraphics*{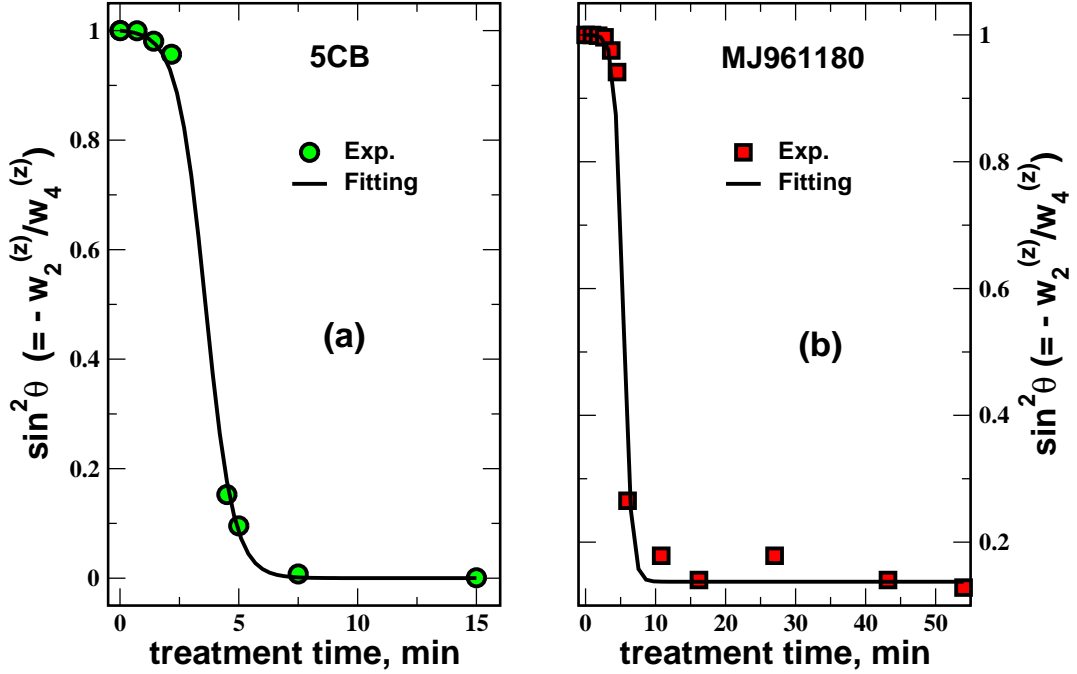}}
\caption{%
Anchoring parameter $-w_2/w_4$ [$=\sin^2\theta$] as a function
of the treatment time measured in (a)~5CB
and (b)~MJ961180 cells.
Experimental data are fitted using
the formula~\eqref{eq:fitting} at $\gamma=1.7$~min$^{-1}$.
The curves shown as solid lines are computed 
with the following dimensionless fitting parameters:
(a)~$\beta_1\approx 2.78\times 10^{-7}$ and
$\beta_2\approx 2.21\times 10^{-3}$;
(b)~$\beta_1\approx 1.52\times 10^{-5}$ and
$\beta_2\approx 1.1\times 10^{-4}$.
}
\label{fig:fitting}
\end{figure*}

Mechanisms behind the azimuthal anchoring
transition in 5CB cells are much less clear. 
Our assumption is 
that it is governed by the topography factor. 
Our previous results~\cite{Yarosh:lc:2004,Yarosh:sid:2005}
suggest that the plasma beam may produce anisotropy of the surface
relief. In addition, microgrooves generated by 
plasma beam from ALS
were recently detected in~\cite{Lin:jjap:2006}.  
With increasing the exposure dose
topographical anisotropy changes direction from the projection of plasma
beam on the substrate to the perpendicular direction. 

This might explain the azimuthal transition from 
the 2nd type alignment to the 3rd type alignment for LC 5CB. 
Such anchoring transition seems to be
possible only if the pretilt angle is sufficiently low,
whereas the topography factor is of minor importance
at high pretilt angles. 



Interestingly, 
our results concerning the materials of different signs of the
dielectric anisotropy bear close similarity to those
reported in Ref.~\cite{Lu:jjap:2004}
where orientation of NLC with positive (E8) and 
negative (MLC 95-465) 
dielectric anisotropy on obliquely evaporated SiO$_2$ films
in relation to the evaporation angle $\alpha$ was studied
experimentally.
For E8, alignment is
planar (and perpendicular to the incidence plane) at
sufficiently small evaporation angles ($\alpha <60$~deg),
whereas, for $\alpha>60$ deg, 
the alignment is tilted in the incidence plane. 
By contrast, for $\Delta\epsilon<0$, 
the director remains in the plane of incidence and
alignment approaches the homeotropic structure as 
the evaporation angle decreases. 
Different alignment behavior of the NLC materials is attributed to the Van der Waals (VdW) interaction
and order electricity. 
The competing effects between the VdW and dipole-to-dipole
interactions are considered in~\cite{Bos:jap:2006}
to explain why vertical alignment of
NLC with negative dielectric anisotropy
on both SiO$_x$ and PI can be improved by doping
with a positive dielectric material such as 5CB.



In closing this section we discuss
a simple phenomenological model that 
can be used to describe both the zenithal and
azimuthal anchoring transitions qualitatively.
Typically, such models are formulated in terms of
the phenomenological expressions for
the anchoring energy potential.
So, the orientational structure in a uniformly aligned NLC cell
is determined by the easy axis, which can be computed
by minimizing the anchoring energy.

First we consider the most extensively studied
case of isotropic flat substrates
where the symmetry of the surface is characterized by its normal,
$\uvc{k}=\uvc{z}$.
So, the anchoring energy can be written as a function
of the pretilt angle, $\theta$, in the following generalized form
\begin{equation}
  \label{eq:gen-Parson}
  W_{\text{P}}(c)=w_{1}\, c +\frac{w_{2}}{2}\,c^2
+\frac{w_{4}}{4}\,c^4,
\end{equation}
where $c\equiv n_z=\sin\theta$ is the $z$ component of 
the NLC director $\vc{n}$.
At $w_{1}=w_{4}=0$, the energy~\eqref{eq:gen-Parson}
simplifies giving the well known Papini-Papoular
potential~\cite{Rap:1969,Gennes:bk:1993}.

The first term on the right hand side of Eq.~\eqref{eq:gen-Parson}
breaks equivalence between $\vc{n}$ and $-\vc{n}$
due to polar ordering effects in the interfacial 
layer~\cite{Parsons:prl:1978,McMullen:pra:1988}.
The model with this polarity breaking term
and $w_4=0$ (the Parson's model) 
was previously employed to describe
anchoring transitions on oxidized silane
substrates~\cite{Hubert:epjb:1999,Fonseca:jap:2003}
and in freely suspended nematic films~\cite{Sonin:pre:1995}.

The expression for the anchoring energy 
with the positive fourth order coefficient $w_4$ coming from the 
quadrupole-quadrupole
interactions and short-range anisotropic repulsive and attractive 
forces~\cite{Sluck:jcp1:1992,Sluck:jpf:1993,Braun:pre:1999}
was originally derived by Sen and Sullivan~\cite{Sen:1987}. 
The non-polar  azimuthally degenerated anchoring
energy~\eqref{eq:gen-Parson} with the quartic term
($w_1=0$ and $w_4\ne 0$) 
was recently used to analyze
temperature-driven transitions between the conical, planar and
anticonical anchorings 
observed on a grafted polymer brush~\cite{Faget:pre:2006}. 

Anchoring properties of the generalized
potential~\eqref{eq:gen-Parson}
can be conveniently characterized by the anchoring phase diagram
in the $w_2$--$w_4$ parameter plane.
We present the phase anchoring diagrams
for two cases: (a)~the non-polar model 
in the Sen-Sullivan form with $w_1=0$
(see Fig.~\ref{fig:nonpolar})
and (b)~the generalized polar model with $w_1\ne0$
(see Fig.~\ref{fig:polar}).

Referring to Fig.~\ref{fig:nonpolar},
when $w_1=0$ and the fourth order (quartic) coefficient $w_4$ is positive,
the regions of homeotropic ($\vc{N}$), 
tilted ($\vc{T}$), and planar anchorings ($\vc{P}$), 
are separated by two  
solid lines, $L_2$ and $w_2=0$,
where the second order transitions take place.
More generally, the symbol $\vc{N}$ ($\vc{P}$) mark regions
where the easy axis is normal (parallel) to 
the specified reference plane such as the plane of substrates
or the incidence plane.

By contrast, if $w_4$ is negative,
the transition between planar and homeotropic structures
is discontinuous and does not involve tilted configurations.
The structures are of the same energy 
at the points on the dashed line $L_1$. 

In the coexistence regions, $\vc{N_B}$ and
$\vc{P_B}$, enclosed by the dash-dotted lines, $L'_2$ and $w_2=0$,
there is an energy barrier between the homeotropic and planar
anchorings. According to Ref.~\cite{Faget:pre:2006},
the latter can be referred to as the anticonical anchoring.
Note that the above results were previously reported
for differently parameterized anchoring potentials 
in~\cite{Sluck:jcp1:1992,Faget:pre:2006}. 

In Fig.~\ref{fig:polar} we show the anchoring diagram for 
the less familiar case of the generalized model
with non-vanishing polar coefficient $w_1$.
The diagram does not depend on the sign of the polar coefficient
because the potential~\eqref{eq:gen-Parson} is invariant
under the symmetry transformation: $c\to -c$ and $w_1\to -w_1$.

For $w_4>-|w_1|/2$, similar to the non-polar model,
the solid line $L_2$ defines the second order transition
between the homeotropic and tilted structures.
Contrastingly, the second order transition between tilted
and planar anchorings is suppressed
as there are no regions of planar anchoring at $w_1\ne 0$. 
>From Fig.~\ref{fig:polar} this transition appears to be 
replaced with crossing 
the boundary curve $L_{\text{d}}$ between the regions
$\vc{T_B}$ and $\vc{T}$
(the dash-dotted line $L_{\text{d}}$ above the point $O_2$).
So, the anchoring in the region $\vc{T}$
characterized by the tilted equilibrium
structure and the absence of metastable states
can be regarded as a counterpart of the planar structure
(region $\vc{P}$ in Fig.~\ref{fig:nonpolar}).


In Fig.~\ref{fig:polar}, the line corresponding to 
the first order transition is depicted as the dashed curve $L_1$.
The latter can be derived in the following parameterized form 
\begin{equation}
  \label{eq:curve-L1}
  L_{1}=
\begin{cases}
w_{2}=|w_1|\,t^{-1}[1+2t^2(1+t)^{-2}],\\
w_{4}=-2|w_1|\, t^{-1} (1+t)^{-2},
\end{cases}
\end{equation}
where the parameter $t$, $0<t\le 1$, defines the tilted configuration,
$c_{\text{tlt}}=-t\,w_1/|w_1|$, which is energetically equivalent
to the homeotropic structure:
$W_{\text{P}}(c_{\text{tlt}})=W_{\text{P}}(c_{\text{hom}})$
at
$c_{\text{hom}}=-w_1/|w_1|$.

It is clear that  both the polar and non-polar models predict
the anchoring transitions that can be either continuous or
discontinuous depending on the value of the fourth order coefficient
$w_4$. It turned out that suppressing the planar anchoring
is one of the most crucial effects induced by 
the polar term proportional to $w_1$.
This effect can also be seen from the curves for the pretilt angle 
presented in Fig.~\ref{fig:pretilt-theor} and
computed as a function of the dimensionless parameter
$\tilde{w_2}\equiv w_2/w_4$ at the fixed ratio $w_1$ and $w_4$.

The model~\eqref{eq:gen-Parson} is azimuthally degenerated
and thus cannot be applied directly to 
the transitions observed in our experiments.
The important point is that the incidence plane of irradiation 
with the normal directed along the $y$ axis
(see Fig.~\ref{fig:geometry}) 
has to be taken into account
as an additional element of the surface geometry.

By the same reasoning as for obliquely evaporated 
SiO$_x$~\cite{Zvezd:pre:2001}
we find, on symmetry grounds, 
that the anchoring potential may additionally depend
on $n_y^2$ and the model~\eqref{eq:gen-Parson}
can be extended as follows
\begin{align}
  \label{eq:anchor-poten}
&
W=W_z(n_z)+W_y(n_y),
\\
&
  W_{a}(n_{a})=\frac{w_{2}^{(a)}}{2}\,n_{a}^2
+\frac{w_{4}^{(a)}}{4}\,n_{a}^4,\quad
a\in\{z,\,y\}.
\notag
\end{align}
Note that the polar anchoring terms are neglected 
in the energy~\eqref{eq:anchor-poten},  so as not to rule out 
experimentally observed planar anchoring and
the structures tilted in the plane of incidence.

Generally, the $n_y$ dependent  terms in the extended model 
arise from the reduction of symmetry caused by anisotropy
of the substrates.
In particular, under certain conditions, the energy~\eqref{eq:anchor-poten}
can be derived from the anchoring potential
obtained  in~\cite{Kis:pre2:2005} for azo-dye photoaligned films.

The structure of the expression~\eqref{eq:anchor-poten}
bears close resemblance to the models 
formulated in terms of two competing anchoring directions (easy axes).
In the Rapini-Papoular approximation,
such dual axis models were previously employed to describe
light-induced anchoring transitions in~\cite{Andrien:jetp:1997}
and to study competitive effects of photoalignment and 
microgrooves in~\cite{Osipov:jap:2002}.
Anchoring properties of rubbed polyimide alignment layers
were also studied by using the model 
supplemented with the fourth order 
term in~\cite{Shio:pre:2003,Huang:apl:2005}.

For the model~\eqref{eq:anchor-poten}, the anchoring transitions
can be geometrically described in terms of two points:
$\vc{w}_z\equiv (w_2^{(z)}, w_4^{(z)})$ and $\vc{w}_y\equiv
(w_2^{(y)}, w_4^{(y)})$, so that
the plane of reference is the substrate and the incidence plane
for $\vc{w}_z$ and $\vc{w}_y$, respectively.
These points both lie in the $w_2$-$w_4$ plane
and dependence of the anchoring coefficients,
$\vc{w}_z$ and $\vc{w}_y$,
on the irradiation dose can be depicted as two trajectories.
The trajectories are illustrated in Fig.~\ref{fig:nonpolar}
under the simplifying assumption that
the fourth order coefficients, $w_4^{(z)}$ and $w_4^{(y)}$,
are kept constant
being independent of the treatment time.

The continuous homeotropic-to-planar transition occurs
when the point $\vc{w}_z$ moves from its initial position
in the region of homeotropic anchoring, $\vc{w}_{z}^{(i)}\in\vc{N}$,
to the final state of planar anchoring with $\vc{w}_{z}^{(f)}\in\vc{P}$
through the region of tilted structures, $\vc{T}$.
Reorientation of the director takes place 
in the incidence ($x$-$z$) plane 
provided $\vc{w}_y$ stay in the region $P$
during the zenithal transition.

If $\vc{w}_z$ is in the region $\vc{P}$,
anchoring is planar and the director orientation is determined
by the position of the point $\vc{w}_y$.
The azimuthal transition between planar structures 
aligned parallel ($n_y=0$) and normal ($n_y^2=1$) 
to the incidence plane can be depicted as the line
connecting two points: $\vc{w}_{y}^{(i)}\in\vc{P}$ and 
$\vc{w}_{y}^{(f)}\in\vc{N}$ (see Fig.~\ref{fig:nonpolar}).

Now we demonstrate that
the zenithal anchoring transition can be described 
quantitatively.
For this purpose we take the assumption of
exponential dependence of
the anchoring coefficients, $w_2^{(z)}$ and $w_4^{(z)}$,
on the exposure time. 

On this assumption, the simplest analytical relation 
for the pretilt angle, $\theta$, can be written in 
the following form:
\begin{equation}
  \label{eq:fitting}
  \sin^2\theta=-\frac{w_2^{(z)}}{w_4^{(z)}}=
\frac{1+\beta_1 [\exp(\gamma\tau_{\text{exp}})-1]}{%
1+\beta_2 [\exp(\gamma\tau_{\text{exp}})-1]},
\end{equation}
where $\tau_{\text{exp}}$ is the exposure (treatment) time.
The results of phenomenological models for
different photo-oriented films~\cite{Chen:1996,Kis:jpcm:2002,Kis:pre:2003} 
and for aligning layers produced by collimated ion 
beams~\cite{Chau:nat:2001} both suggest that 
the exponential dependence is typical for the corresponding 
concentrations. So, in our case, it  can be regarded as a reasonable 
approximation for the concentration of hydrophobic chains.
Note that, strictly speaking, computing the pretilt angle
requires a rather involved theoretical analysis 
which is beyond the scope of this paper. 

The expression~\eqref{eq:fitting} can be used to fit the experimental
data for 5CB and MJ961180 cells. The results of calculations are
presented in Fig.~\ref{fig:fitting}. Clearly, they show that
the difference between the materials is determined by
the two fitting parameters: $\beta_1$ and $\beta_2$.
In particular, for 5CB cells, the ratio $\beta_1/\beta_2$ 
appears to be negligibly small and, as a result, the fourth order
coefficient $w_4^{(z)}$ is almost independent of the irradiation dose.
But this is not the case for MJ961180 cells.

So, the experimentally observed transitions
can be modeled by using the phenomenological anchoring 
potential~\eqref{eq:anchor-poten}.
Note that, in the strict sense, our experiments do not imply 
the polar anchoring terms 
proportional to $w_1^{(z)}$ and $w_1^{(y)}$ are identically absent
for all exposure doses.
We can only conclude that the coefficient $w_1^{(z)}$ vanishes
for 5CB cells at high irradiation doses in the region of planar
anchorings, whereas the coefficient $w_1^{(y)}$ is zero 
during reorientation in the plane of incidence.

Interestingly, the model~\eqref{eq:anchor-poten} can also be applied to 
the temperature induced anchoring transition on 
a SiO$_x$ surface~\cite{Zvezd:pre:2001}.
 It can be shown that,
when
$w_2^{(y)}$ varies from $-w_4^{(y)}$ to zero lying
on the line
\begin{equation}
  \label{eq:SiO-line}
  \frac{w_2^{(z)}}{w_4^{(z)}}+
\sin^{2}\alpha\,
\left[
  \frac{w_2^{(y)}}{w_4^{(y)}} +1
\right]=0,
\end{equation}
anchoring changes from planar, 
$\vc{n}=(0,1,0)$,  to tilted, $\vc{n}=(\cos\alpha,0,\sin\alpha)$, 
with the director moving on the plane 
that forms the angle $\alpha$ with the film.
Qualitatively, this reproduces behavior of the NLC director 
in the course of the mixed anchoring transitions on
obliquely evaporated SiO$_x$ films.


\section{Conclusions}
\label{sec:concl}
 
We have observed experimentally
the second order zenithal anchoring transitions in liquid crystals
with positive and negative dielectric anisotropy
oriented by hydrophobic substrates obliquely 
processed with a plasma beam. 
 The transition is characterized by a pronounced decline of 
the pretilt angle with the exposure dose
upon reaching  the critical value of
surface free energy related to 
the critical concentration of hydrophobic
chains at the surface. 

In LC 5CB with $\Delta\epsilon>0$,
the zenithal transition is followed by
the azimuthal transition when
the exposure dose increases further. 
It occurs through the in-plane reorientation of the easy axis
which is rotated by a right angle. 
This reorientation  is found to involve 
two-fold degenerated transient structures
and, as a consequence, we arrive at the conclusion that 
the azimuthal transition is second order.
This transition can be reasonably explained  
by experimentally detected change of 
topographical anisotropy 

We have formulated a simple phenomenological model
where two competing anchoring directions appear
as a result of additional plasma beam induced anisotropy
of the treated substrate.
In order to perform qualitative analysis of this model,
the anchoring diagrams of the generalized potential 
were studied for both polar and non-polar cases. 
The result is that the experimentally observed 
anchoring transitions can be properly modeled using  
the non-polar dual axis model supplemented with
the fourth order terms.

In conclusion, it should be noted that all types of LC alignment
observed in our experiments such as 
high and low pretilt structures along with planar alignment 
are of considerable interest for applications. 
The technology related issues were briefly discussed in our previous
publications~\cite{Yarosh:lc:2004,Yarosh:sid:2005}.

\begin{acknowledgments}
This work was performed under INTAS Grant No.~03-51-5448.
O.V.Ya and R.M.K acknowledge financial support from NASU
under grant  No.~10/07-H-32.
We also thank Dr. I.~Gerus 
(Institute of Petrol and Biochemistry of NASU, Kyiv, Ukraine)
for providing us with the fluorinated polyimide.
 \end{acknowledgments}
 

\begin{thebibliography}{53}
\expandafter\ifx\csname natexlab\endcsname\relax\def\natexlab#1{#1}\fi
\expandafter\ifx\csname bibnamefont\endcsname\relax
  \def\bibnamefont#1{#1}\fi
\expandafter\ifx\csname bibfnamefont\endcsname\relax
  \def\bibfnamefont#1{#1}\fi
\expandafter\ifx\csname citenamefont\endcsname\relax
  \def\citenamefont#1{#1}\fi
\expandafter\ifx\csname url\endcsname\relax
  \def\url#1{\texttt{#1}}\fi
\expandafter\ifx\csname urlprefix\endcsname\relax\def\urlprefix{URL }\fi
\providecommand{\bibinfo}[2]{#2}
\providecommand{\eprint}[2][]{\url{#2}}

\bibitem[{\citenamefont{Sluckin and Poniewierski}(1986)}]{Sluck:in:1986}
\bibinfo{author}{\bibfnamefont{T.~J.} \bibnamefont{Sluckin}} \bibnamefont{and}
  \bibinfo{author}{\bibfnamefont{A.}~\bibnamefont{Poniewierski}}, in
  \emph{\bibinfo{booktitle}{Fluid Interfacial Phenomena}}, edited by
  \bibinfo{editor}{\bibfnamefont{C.~A.} \bibnamefont{Croxton}}
  (\bibinfo{publisher}{Wiley}, \bibinfo{address}{Chichester},
  \bibinfo{year}{1986}), chap.~\bibinfo{chapter}{5}, pp.
  \bibinfo{pages}{215--253}.

\bibitem[{\citenamefont{J\'er\^ome}(1991)}]{Jerom:rpp:1991}
\bibinfo{author}{\bibfnamefont{B.}~\bibnamefont{J\'er\^ome}},
  \bibinfo{journal}{Rep. Prog. Phys.} \textbf{\bibinfo{volume}{54}},
  \bibinfo{pages}{391} (\bibinfo{year}{1991}).

\bibitem[{\citenamefont{Barbero and Durand}(1996)}]{Durand:1996}
\bibinfo{author}{\bibfnamefont{G.}~\bibnamefont{Barbero}} \bibnamefont{and}
  \bibinfo{author}{\bibfnamefont{G.}~\bibnamefont{Durand}}, in
  \emph{\bibinfo{booktitle}{Liquid Crystals in Complex Geometries}}, edited by
  \bibinfo{editor}{\bibfnamefont{G.~P.} \bibnamefont{Crawford}}
  \bibnamefont{and} \bibinfo{editor}{\bibfnamefont{S.}~\bibnamefont{\v{Z}umer}}
  (\bibinfo{publisher}{Taylor \& Francis}, \bibinfo{address}{London},
  \bibinfo{year}{1996}), chap.~\bibinfo{chapter}{2}, pp.
  \bibinfo{pages}{21--52}.

\bibitem[{\citenamefont{K\"{a}nel et~al.}(1981)\citenamefont{K\"{a}nel,
  Litster, Melngailis, and Smith}}]{Kanel:pra:1981}
\bibinfo{author}{\bibfnamefont{H.~V.} \bibnamefont{K\"{a}nel}},
  \bibinfo{author}{\bibfnamefont{J.~D.} \bibnamefont{Litster}},
  \bibinfo{author}{\bibfnamefont{J.}~\bibnamefont{Melngailis}},
  \bibnamefont{and} \bibinfo{author}{\bibfnamefont{H.~I.} \bibnamefont{Smith}},
  \bibinfo{journal}{Phys. Rev. A} \textbf{\bibinfo{volume}{24}},
  \bibinfo{pages}{2713} (\bibinfo{year}{1981}).

\bibitem[{\citenamefont{Alkhairalla et~al.}(1999)\citenamefont{Alkhairalla,
  Allinson, Boden, Evans, and Henderson}}]{Alkhair:pre:1999}
\bibinfo{author}{\bibfnamefont{B.}~\bibnamefont{Alkhairalla}},
  \bibinfo{author}{\bibfnamefont{H.}~\bibnamefont{Allinson}},
  \bibinfo{author}{\bibfnamefont{N.}~\bibnamefont{Boden}},
  \bibinfo{author}{\bibfnamefont{S.~D.} \bibnamefont{Evans}}, \bibnamefont{and}
  \bibinfo{author}{\bibfnamefont{J.~R.} \bibnamefont{Henderson}},
  \bibinfo{journal}{Phys. Rev. E} \textbf{\bibinfo{volume}{59}},
  \bibinfo{pages}{3033} (\bibinfo{year}{1999}).

\bibitem[{\citenamefont{Chiarelli et~al.}(1983)\citenamefont{Chiarelli, Faetti,
  and Fronzoni}}]{Chiarelli:jpf:1983}
\bibinfo{author}{\bibfnamefont{P.}~\bibnamefont{Chiarelli}},
  \bibinfo{author}{\bibfnamefont{S.}~\bibnamefont{Faetti}}, \bibnamefont{and}
  \bibinfo{author}{\bibfnamefont{L.}~\bibnamefont{Fronzoni}},
  \bibinfo{journal}{J. Physique} \textbf{\bibinfo{volume}{44}},
  \bibinfo{pages}{1061} (\bibinfo{year}{1983}).

\bibitem[{\citenamefont{Chiarelli et~al.}(1984)\citenamefont{Chiarelli, Faetti,
  and Fronzoni}}]{Chiarelli:pla:1984}
\bibinfo{author}{\bibfnamefont{P.}~\bibnamefont{Chiarelli}},
  \bibinfo{author}{\bibfnamefont{S.}~\bibnamefont{Faetti}}, \bibnamefont{and}
  \bibinfo{author}{\bibfnamefont{L.}~\bibnamefont{Fronzoni}},
  \bibinfo{journal}{Phys. Lett. A} \textbf{\bibinfo{volume}{101}},
  \bibinfo{pages}{31} (\bibinfo{year}{1984}).

\bibitem[{\citenamefont{Sonin et~al.}(1995)\citenamefont{Sonin, Yethiraj,
  Bechhoefer, and Frisken}}]{Sonin:pre:1995}
\bibinfo{author}{\bibfnamefont{A.~A.} \bibnamefont{Sonin}},
  \bibinfo{author}{\bibfnamefont{A.}~\bibnamefont{Yethiraj}},
  \bibinfo{author}{\bibfnamefont{J.}~\bibnamefont{Bechhoefer}},
  \bibnamefont{and} \bibinfo{author}{\bibfnamefont{B.~J.}
  \bibnamefont{Frisken}}, \bibinfo{journal}{Phys. Rev. E}
  \textbf{\bibinfo{volume}{52}}, \bibinfo{pages}{6260} (\bibinfo{year}{1995}).

\bibitem[{\citenamefont{Shioda et~al.}(2003)\citenamefont{Shioda, Wen, and
  Rosenblatt}}]{Shio:pre:2003}
\bibinfo{author}{\bibfnamefont{T.}~\bibnamefont{Shioda}},
  \bibinfo{author}{\bibfnamefont{B.}~\bibnamefont{Wen}}, \bibnamefont{and}
  \bibinfo{author}{\bibfnamefont{C.}~\bibnamefont{Rosenblatt}},
  \bibinfo{journal}{Phys. Rev. E} \textbf{\bibinfo{volume}{67}},
  \bibinfo{pages}{041706} (\bibinfo{year}{2003}).

\bibitem[{\citenamefont{Pieranski and J\'er\^ome}(1989)}]{Jerome:pra:1989}
\bibinfo{author}{\bibfnamefont{P.}~\bibnamefont{Pieranski}} \bibnamefont{and}
  \bibinfo{author}{\bibfnamefont{B.}~\bibnamefont{J\'er\^ome}},
  \bibinfo{journal}{Phys. Rev. A} \textbf{\bibinfo{volume}{40}},
  \bibinfo{pages}{317} (\bibinfo{year}{1989}).

\bibitem[{\citenamefont{Bechhoefer
  et~al.}(1990{\natexlab{a}})\citenamefont{Bechhoefer, J\'er\^ome, and
  Pieranski}}]{Jerome:pra:1990}
\bibinfo{author}{\bibfnamefont{J.}~\bibnamefont{Bechhoefer}},
  \bibinfo{author}{\bibfnamefont{B.}~\bibnamefont{J\'er\^ome}},
  \bibnamefont{and}
  \bibinfo{author}{\bibfnamefont{P.}~\bibnamefont{Pieranski}},
  \bibinfo{journal}{Phys. Rev. A} \textbf{\bibinfo{volume}{41}},
  \bibinfo{pages}{3187} (\bibinfo{year}{1990}{\natexlab{a}}).

\bibitem[{\citenamefont{Bechhoefer
  et~al.}(1990{\natexlab{b}})\citenamefont{Bechhoefer, Duvai, Masson,
  J\'er\^ome, Hornreich, and Pieranski}}]{Jerome:prl:1990}
\bibinfo{author}{\bibfnamefont{J.}~\bibnamefont{Bechhoefer}},
  \bibinfo{author}{\bibfnamefont{J.-L.} \bibnamefont{Duvai}},
  \bibinfo{author}{\bibfnamefont{L.}~\bibnamefont{Masson}},
  \bibinfo{author}{\bibfnamefont{B.}~\bibnamefont{J\'er\^ome}},
  \bibinfo{author}{\bibfnamefont{R.~M.} \bibnamefont{Hornreich}},
  \bibnamefont{and}
  \bibinfo{author}{\bibfnamefont{P.}~\bibnamefont{Pieranski}},
  \bibinfo{journal}{Phys. Rev. Lett.} \textbf{\bibinfo{volume}{64}},
  \bibinfo{pages}{1911} (\bibinfo{year}{1990}{\natexlab{b}}).

\bibitem[{\citenamefont{Bechhoefer et~al.}(1991)\citenamefont{Bechhoefer,
  J\'er\^ome, and Pieranski}}]{Jerome:phtr:1991}
\bibinfo{author}{\bibfnamefont{J.}~\bibnamefont{Bechhoefer}},
  \bibinfo{author}{\bibfnamefont{B.}~\bibnamefont{J\'er\^ome}},
  \bibnamefont{and}
  \bibinfo{author}{\bibfnamefont{P.}~\bibnamefont{Pieranski}},
  \bibinfo{journal}{Phase Transitions} \textbf{\bibinfo{volume}{33}},
  \bibinfo{pages}{227} (\bibinfo{year}{1991}).

\bibitem[{\citenamefont{Gibbons et~al.}(1991)\citenamefont{Gibbons, Shannon,
  Sun, and Swetlin}}]{Gibbon:nat:1991}
\bibinfo{author}{\bibfnamefont{W.~M.} \bibnamefont{Gibbons}},
  \bibinfo{author}{\bibfnamefont{P.~J.} \bibnamefont{Shannon}},
  \bibinfo{author}{\bibfnamefont{S.-T.} \bibnamefont{Sun}}, \bibnamefont{and}
  \bibinfo{author}{\bibfnamefont{B.~J.} \bibnamefont{Swetlin}},
  \bibinfo{journal}{Nature} \textbf{\bibinfo{volume}{351}}, \bibinfo{pages}{49}
  (\bibinfo{year}{1991}).

\bibitem[{\citenamefont{Li}(1995)}]{Li:lc:1995}
\bibinfo{author}{\bibfnamefont{Z.}~\bibnamefont{Li}}, \bibinfo{journal}{Liq.
  Cryst.} \textbf{\bibinfo{volume}{19}}, \bibinfo{pages}{307}
  (\bibinfo{year}{1995}).

\bibitem[{\citenamefont{Andrienko et~al.}(1997)\citenamefont{Andrienko, Kurioz,
  Reznikov, and Reshetnyak}}]{Andrien:jetp:1997}
\bibinfo{author}{\bibfnamefont{D.}~\bibnamefont{Andrienko}},
  \bibinfo{author}{\bibfnamefont{Y.}~\bibnamefont{Kurioz}},
  \bibinfo{author}{\bibfnamefont{Y.}~\bibnamefont{Reznikov}}, \bibnamefont{and}
  \bibinfo{author}{\bibfnamefont{Y.}~\bibnamefont{Reshetnyak}},
  \bibinfo{journal}{JETP} \textbf{\bibinfo{volume}{112}}, \bibinfo{pages}{2045}
  (\bibinfo{year}{1997});
\bibinfo{author}{\bibfnamefont{Y.~A.} \bibnamefont{Reznikov}} \bibnamefont{and}
  \bibinfo{author}{\bibfnamefont{O.~V.} \bibnamefont{Yaroshchuk}}, in
  \emph{\bibinfo{booktitle}{Abstracts of OLC'95}} (\bibinfo{publisher}{VI
  Intern. Topical Meeting on Opt. of Liq. Cryst.}, \bibinfo{address}{Le
  Touquet, France}, \bibinfo{year}{1995}), p.~\bibinfo{pages}{42}.

\bibitem[{\citenamefont{Andrienko et~al.}(1998)\citenamefont{Andrienko,
  Dyadyusha, Kurioz, Reshetnyak, and Reznikov}}]{Andrienko:mclc:1998}
\bibinfo{author}{\bibfnamefont{D.}~\bibnamefont{Andrienko}},
  \bibinfo{author}{\bibfnamefont{A.}~\bibnamefont{Dyadyusha}},
  \bibinfo{author}{\bibfnamefont{Y.}~\bibnamefont{Kurioz}},
  \bibinfo{author}{\bibfnamefont{V.}~\bibnamefont{Reshetnyak}},
  \bibnamefont{and} \bibinfo{author}{\bibfnamefont{Y.}~\bibnamefont{Reznikov}},
  \bibinfo{journal}{Mol. Cryst. Liq. Cryst.} \textbf{\bibinfo{volume}{321}},
  \bibinfo{pages}{299} (\bibinfo{year}{1998}).

\bibitem[{\citenamefont{Komitov et~al.}(2000)\citenamefont{Komitov, Ichimura,
  and Strigazzi}}]{Stri:2000}
\bibinfo{author}{\bibfnamefont{L.}~\bibnamefont{Komitov}},
  \bibinfo{author}{\bibfnamefont{K.}~\bibnamefont{Ichimura}}, \bibnamefont{and}
  \bibinfo{author}{\bibfnamefont{A.}~\bibnamefont{Strigazzi}},
  \bibinfo{journal}{Liq. Cryst.} \textbf{\bibinfo{volume}{27}},
  \bibinfo{pages}{51} (\bibinfo{year}{2000}).

\bibitem[{\citenamefont{O'Neill and Kelly}(2000)}]{Kelly:jpd:2000}
\bibinfo{author}{\bibfnamefont{M.}~\bibnamefont{O'Neill}} \bibnamefont{and}
  \bibinfo{author}{\bibfnamefont{S.~M.} \bibnamefont{Kelly}},
  \bibinfo{journal}{J. Phys. D: Appl. Phys.} \textbf{\bibinfo{volume}{33}},
  \bibinfo{pages}{R67} (\bibinfo{year}{2000}).

\bibitem[{\citenamefont{Chigrinov et~al.}(2003)\citenamefont{Chigrinov,
  Kozenkov, and Kwok}}]{Chigr:rewiev:2003}
\bibinfo{author}{\bibfnamefont{V.~G.} \bibnamefont{Chigrinov}},
  \bibinfo{author}{\bibfnamefont{V.~M.} \bibnamefont{Kozenkov}},
  \bibnamefont{and} \bibinfo{author}{\bibfnamefont{H.~S.} \bibnamefont{Kwok}},
  in \emph{\bibinfo{booktitle}{Optical applications in photoaligning}}, edited
  by \bibinfo{editor}{\bibfnamefont{L.}~\bibnamefont{Vicari}}
  (\bibinfo{publisher}{Inst. of Physics}, \bibinfo{address}{Bristol, UK},
  \bibinfo{year}{2003}), pp. \bibinfo{pages}{201--244}.

\bibitem[{\citenamefont{Janning}(1972)}]{Jannin:apl:1972}
\bibinfo{author}{\bibfnamefont{J.~L.} \bibnamefont{Janning}},
  \bibinfo{journal}{Appl. Phys. Lett.} \textbf{\bibinfo{volume}{21}},
  \bibinfo{pages}{173} (\bibinfo{year}{1972}).

\bibitem[{\citenamefont{Urbach et~al.}(1974)\citenamefont{Urbach, Boix, and
  Guyon}}]{Urbach:apl:1974}
\bibinfo{author}{\bibfnamefont{W.}~\bibnamefont{Urbach}},
  \bibinfo{author}{\bibfnamefont{M.}~\bibnamefont{Boix}}, \bibnamefont{and}
  \bibinfo{author}{\bibfnamefont{E.}~\bibnamefont{Guyon}},
  \bibinfo{journal}{Appl. Phys. Lett.} \textbf{\bibinfo{volume}{25}},
  \bibinfo{pages}{479} (\bibinfo{year}{1974}).

\bibitem[{\citenamefont{Jerome et~al.}(1988)\citenamefont{Jerome, Boix, and
  Pieranski}}]{Jerome:eurpl:1988}
\bibinfo{author}{\bibfnamefont{B.}~\bibnamefont{Jerome}},
  \bibinfo{author}{\bibfnamefont{M.}~\bibnamefont{Boix}}, \bibnamefont{and}
  \bibinfo{author}{\bibfnamefont{P.}~\bibnamefont{Pieranski}},
  \bibinfo{journal}{Eurphys. Lett.} \textbf{\bibinfo{volume}{5}},
  \bibinfo{pages}{693} (\bibinfo{year}{1988}).

\bibitem[{\citenamefont{Monkade et~al.}(1988)\citenamefont{Monkade, Boix, and
  Durand}}]{Monkade:eurpl:1988}
\bibinfo{author}{\bibfnamefont{M.}~\bibnamefont{Monkade}},
  \bibinfo{author}{\bibfnamefont{M.}~\bibnamefont{Boix}}, \bibnamefont{and}
  \bibinfo{author}{\bibfnamefont{G.}~\bibnamefont{Durand}},
  \bibinfo{journal}{Eurphys. Lett.} \textbf{\bibinfo{volume}{5}},
  \bibinfo{pages}{697} (\bibinfo{year}{1988}).

\bibitem[{\citenamefont{Chaudhari et~al.}(2001)\citenamefont{Chaudhari, Lacey,
  Doyle, Galligan, Lien, Callegari, Hougham, Lang, Andry, John
  et~al.}}]{Chau:nat:2001}
\bibinfo{author}{\bibfnamefont{P.}~\bibnamefont{Chaudhari}},
  \bibinfo{author}{\bibfnamefont{J.}~\bibnamefont{Lacey}},
  \bibinfo{author}{\bibfnamefont{J.}~\bibnamefont{Doyle}},
  \bibinfo{author}{\bibfnamefont{E.}~\bibnamefont{Galligan}},
  \bibinfo{author}{\bibfnamefont{S.-H.~A.} \bibnamefont{Lien}},
  \bibinfo{author}{\bibfnamefont{A.}~\bibnamefont{Callegari}},
  \bibinfo{author}{\bibfnamefont{G.}~\bibnamefont{Hougham}},
  \bibinfo{author}{\bibfnamefont{N.~D.} \bibnamefont{Lang}},
  \bibinfo{author}{\bibfnamefont{P.~S.} \bibnamefont{Andry}},
  \bibinfo{author}{\bibfnamefont{R.}~\bibnamefont{John}}, \bibnamefont{et~al.},
  \bibinfo{journal}{Nature} \textbf{\bibinfo{volume}{411}}, \bibinfo{pages}{56}
  (\bibinfo{year}{2001}).

\bibitem[{\citenamefont{Yaroshchuk et~al.}(2004)\citenamefont{Yaroshchuk,
  Kravchuk, Dobrovolskyy, Qui, and Lavrentovich}}]{Yarosh:lc:2004}
\bibinfo{author}{\bibfnamefont{O.}~\bibnamefont{Yaroshchuk}},
  \bibinfo{author}{\bibfnamefont{R.}~\bibnamefont{Kravchuk}},
  \bibinfo{author}{\bibfnamefont{A.}~\bibnamefont{Dobrovolskyy}},
  \bibinfo{author}{\bibfnamefont{L.}~\bibnamefont{Qui}}, \bibnamefont{and}
  \bibinfo{author}{\bibfnamefont{O.~D.} \bibnamefont{Lavrentovich}},
  \bibinfo{journal}{Liq. Cryst.} \textbf{\bibinfo{volume}{31}},
  \bibinfo{pages}{859} (\bibinfo{year}{2004}).

\bibitem[{\citenamefont{Yaroshchuk et~al.}(2005)\citenamefont{Yaroshchuk,
  Kravchuk, Dobrovolskyy, Liu, and Lee}}]{Yarosh:sid:2005}
\bibinfo{author}{\bibfnamefont{O.}~\bibnamefont{Yaroshchuk}},
  \bibinfo{author}{\bibfnamefont{R.}~\bibnamefont{Kravchuk}},
  \bibinfo{author}{\bibfnamefont{A.}~\bibnamefont{Dobrovolskyy}},
  \bibinfo{author}{\bibfnamefont{P.-C.} \bibnamefont{Liu}}, \bibnamefont{and}
  \bibinfo{author}{\bibfnamefont{C.-D.} \bibnamefont{Lee}},
  \bibinfo{journal}{Journal of SID} \textbf{\bibinfo{volume}{13/4}},
  \bibinfo{pages}{289} (\bibinfo{year}{2005}).

\bibitem[{\citenamefont{Sen and Sullivan}(1987)}]{Sen:1987}
\bibinfo{author}{\bibfnamefont{A.~K.} \bibnamefont{Sen}} \bibnamefont{and}
  \bibinfo{author}{\bibfnamefont{D.~E.} \bibnamefont{Sullivan}},
  \bibinfo{journal}{Phys. Rev. A} \textbf{\bibinfo{volume}{35}},
  \bibinfo{pages}{1391} (\bibinfo{year}{1987}).

\bibitem[{\citenamefont{Zhurin et~al.}(1999)\citenamefont{Zhurin, Kaufman, and
  Robinson}}]{Zhurin:psst:1999}
\bibinfo{author}{\bibfnamefont{V.}~\bibnamefont{Zhurin}},
  \bibinfo{author}{\bibfnamefont{H.}~\bibnamefont{Kaufman}}, \bibnamefont{and}
  \bibinfo{author}{\bibfnamefont{R.}~\bibnamefont{Robinson}},
  \bibinfo{journal}{Plasma Sources Sci. Technol.} \textbf{\bibinfo{volume}{8}},
  \bibinfo{pages}{R1} (\bibinfo{year}{1999}).

\bibitem[{\citenamefont{Hwang et~al.}(2002)\citenamefont{Hwang, Jo, Seo, Rho,
  Lee, and Baik}}]{Hwang:jjap:2002}
\bibinfo{author}{\bibfnamefont{B.~H.} \bibnamefont{Hwang}},
  \bibinfo{author}{\bibfnamefont{Y.~M.} \bibnamefont{Jo}},
  \bibinfo{author}{\bibfnamefont{D.~S.} \bibnamefont{Seo}},
  \bibinfo{author}{\bibfnamefont{S.~J.} \bibnamefont{Rho}},
  \bibinfo{author}{\bibfnamefont{D.~K.} \bibnamefont{Lee}}, \bibnamefont{and}
  \bibinfo{author}{\bibfnamefont{H.~K.} \bibnamefont{Baik}},
  \bibinfo{journal}{Jpn. J. Appl. Phys.} \textbf{\bibinfo{volume}{41}},
  \bibinfo{pages}{L654} (\bibinfo{year}{2002}).

\bibitem[{\citenamefont{Hubert and Galerne}(1999)}]{Hubert:epjb:1999}
\bibinfo{author}{\bibfnamefont{P.}~\bibnamefont{Hubert}} \bibnamefont{and}
  \bibinfo{author}{\bibfnamefont{Y.}~\bibnamefont{Galerne}},
  \bibinfo{journal}{Eur. Phys. J. B} \textbf{\bibinfo{volume}{8}},
  \bibinfo{pages}{245} (\bibinfo{year}{1999}).

\bibitem[{\citenamefont{Fonseca et~al.}(2003)\citenamefont{Fonseca, Hommet, and
  Galerne}}]{Fonseca:jap:2003}
\bibinfo{author}{\bibfnamefont{J.~G.} \bibnamefont{Fonseca}},
  \bibinfo{author}{\bibfnamefont{J.}~\bibnamefont{Hommet}}, \bibnamefont{and}
  \bibinfo{author}{\bibfnamefont{Y.}~\bibnamefont{Galerne}},
  \bibinfo{journal}{Appl. Phys. Lett.} \textbf{\bibinfo{volume}{82}},
  \bibinfo{pages}{58} (\bibinfo{year}{2003}).

\bibitem[{\citenamefont{Jang et~al.}(2006)\citenamefont{Jang, Song, and
  Lee}}]{Jang:jjap:2006}
\bibinfo{author}{\bibfnamefont{E.}~\bibnamefont{Jang}},
  \bibinfo{author}{\bibfnamefont{H.}~\bibnamefont{Song}}, \bibnamefont{and}
  \bibinfo{author}{\bibfnamefont{S.-D.} \bibnamefont{Lee}},
  \bibinfo{journal}{Jpn. J. Appl. Phys.} \textbf{\bibinfo{volume}{45}},
  \bibinfo{pages}{L1238} (\bibinfo{year}{2006}).

\bibitem[{\citenamefont{Akiyama and Iimura}(2001)}]{Akiyama:jjap:2001}
\bibinfo{author}{\bibfnamefont{H.}~\bibnamefont{Akiyama}} \bibnamefont{and}
  \bibinfo{author}{\bibfnamefont{Y.}~\bibnamefont{Iimura}},
  \bibinfo{journal}{Jpn. J. Appl. Phys.} \textbf{\bibinfo{volume}{40}},
  \bibinfo{pages}{L765} (\bibinfo{year}{2001}).

\bibitem[{\citenamefont{Lin and Lee}(2006)}]{Lin:jjap:2006}
\bibinfo{author}{\bibfnamefont{S.-S.} \bibnamefont{Lin}} \bibnamefont{and}
  \bibinfo{author}{\bibfnamefont{Y.-D.} \bibnamefont{Lee}},
  \bibinfo{journal}{Jpn. J. Appl. Phys.} \textbf{\bibinfo{volume}{45}},
  \bibinfo{pages}{L708} (\bibinfo{year}{2006}).

\bibitem[{\citenamefont{Lu}(2006)}]{Lu:jjap:2004}
\bibinfo{author}{\bibfnamefont{M.}~\bibnamefont{Lu}}, \bibinfo{journal}{Jpn. J.
  Appl. Phys.} \textbf{\bibinfo{volume}{43}}, \bibinfo{pages}{8156}
  (\bibinfo{year}{2006}).

\bibitem[{\citenamefont{Chen et~al.}(2006)\citenamefont{Chen, Bos, Kim, and
  Li}}]{Bos:jap:2006}
\bibinfo{author}{\bibfnamefont{C.}~\bibnamefont{Chen}},
  \bibinfo{author}{\bibfnamefont{P.~J.} \bibnamefont{Bos}},
  \bibinfo{author}{\bibfnamefont{J.}~\bibnamefont{Kim}}, \bibnamefont{and}
  \bibinfo{author}{\bibfnamefont{Q.}~\bibnamefont{Li}}, \bibinfo{journal}{J.
  Appl. Phys.} \textbf{\bibinfo{volume}{99}}, \bibinfo{pages}{123523}
  (\bibinfo{year}{2006}).

\bibitem[{\citenamefont{Rapini and Papoular}(1969)}]{Rap:1969}
\bibinfo{author}{\bibfnamefont{A.}~\bibnamefont{Rapini}} \bibnamefont{and}
  \bibinfo{author}{\bibfnamefont{M.}~\bibnamefont{Papoular}},
  \bibinfo{journal}{J. Phys. (Paris) Colloq. C4} \textbf{\bibinfo{volume}{30}},
  \bibinfo{pages}{54} (\bibinfo{year}{1969}).

\bibitem[{\citenamefont{de~Gennes and Prost}(1993)}]{Gennes:bk:1993}
\bibinfo{author}{\bibfnamefont{P.~G.} \bibnamefont{de~Gennes}}
  \bibnamefont{and} \bibinfo{author}{\bibfnamefont{J.}~\bibnamefont{Prost}},
  \emph{\bibinfo{title}{The Physics of Liquid Crystals}}
  (\bibinfo{publisher}{Clarendon Press}, \bibinfo{address}{Oxford},
  \bibinfo{year}{1993}).

\bibitem[{\citenamefont{Parsons}(1978)}]{Parsons:prl:1978}
\bibinfo{author}{\bibfnamefont{J.~D.} \bibnamefont{Parsons}},
  \bibinfo{journal}{Phys. Rev. Lett.} \textbf{\bibinfo{volume}{41}},
  \bibinfo{pages}{877} (\bibinfo{year}{1978}).

\bibitem[{\citenamefont{McMullen}(1988)}]{McMullen:pra:1988}
\bibinfo{author}{\bibfnamefont{W.~E.} \bibnamefont{McMullen}},
  \bibinfo{journal}{Phys. Rev. A} \textbf{\bibinfo{volume}{38}},
  \bibinfo{pages}{6384} (\bibinfo{year}{1988}).

\bibitem[{\citenamefont{Teixeira and Sluckin}(1992)}]{Sluck:jcp1:1992}
\bibinfo{author}{\bibfnamefont{P.~I.~C.} \bibnamefont{Teixeira}}
  \bibnamefont{and} \bibinfo{author}{\bibfnamefont{T.~J.}
  \bibnamefont{Sluckin}}, \bibinfo{journal}{J. Chem. Phys.}
  \textbf{\bibinfo{volume}{97}}, \bibinfo{pages}{1490} (\bibinfo{year}{1992}).

\bibitem[{\citenamefont{Osipov and Sluckin}(1993)}]{Sluck:jpf:1993}
\bibinfo{author}{\bibfnamefont{M.~A.} \bibnamefont{Osipov}} \bibnamefont{and}
  \bibinfo{author}{\bibfnamefont{T.~J.} \bibnamefont{Sluckin}},
  \bibinfo{journal}{J. Phys. II France} \textbf{\bibinfo{volume}{3}},
  \bibinfo{pages}{793} (\bibinfo{year}{1993}).

\bibitem[{\citenamefont{Braun et~al.}(1999)\citenamefont{Braun, Sluckin,
  Velasco, and Mederos}}]{Braun:pre:1999}
\bibinfo{author}{\bibfnamefont{F.~N.} \bibnamefont{Braun}},
  \bibinfo{author}{\bibfnamefont{T.~J.} \bibnamefont{Sluckin}},
  \bibinfo{author}{\bibfnamefont{E.}~\bibnamefont{Velasco}}, \bibnamefont{and}
  \bibinfo{author}{\bibfnamefont{L.}~\bibnamefont{Mederos}},
  \bibinfo{journal}{Phys. Rev. E} \textbf{\bibinfo{volume}{53}},
  \bibinfo{pages}{706} (\bibinfo{year}{1999}).

\bibitem[{\citenamefont{Faget et~al.}(2006)\citenamefont{Faget,
  Lamarque-Forget, Martinot-Lagarde, Auroy, and Dozov}}]{Faget:pre:2006}
\bibinfo{author}{\bibfnamefont{L.}~\bibnamefont{Faget}},
  \bibinfo{author}{\bibfnamefont{S.}~\bibnamefont{Lamarque-Forget}},
  \bibinfo{author}{\bibfnamefont{P.}~\bibnamefont{Martinot-Lagarde}},
  \bibinfo{author}{\bibfnamefont{P.}~\bibnamefont{Auroy}}, \bibnamefont{and}
  \bibinfo{author}{\bibfnamefont{I.}~\bibnamefont{Dozov}},
  \bibinfo{journal}{Phys. Rev. E} \textbf{\bibinfo{volume}{74}},
  \bibinfo{pages}{050701(R)} (\bibinfo{year}{2006}).

\bibitem[{\citenamefont{Barbero et~al.}(2001)\citenamefont{Barbero,
  J\"{a}gemalm, and Zvezdin}}]{Zvezd:pre:2001}
\bibinfo{author}{\bibfnamefont{G.}~\bibnamefont{Barbero}},
  \bibinfo{author}{\bibfnamefont{P.}~\bibnamefont{J\"{a}gemalm}},
  \bibnamefont{and} \bibinfo{author}{\bibfnamefont{A.~K.}
  \bibnamefont{Zvezdin}}, \bibinfo{journal}{Phys. Rev. E}
  \textbf{\bibinfo{volume}{64}}, \bibinfo{pages}{021703}
  (\bibinfo{year}{2001}).

\bibitem[{\citenamefont{Kiselev et~al.}(2005)\citenamefont{Kiselev, Chigrinov,
  and Huang}}]{Kis:pre2:2005}
\bibinfo{author}{\bibfnamefont{A.~D.} \bibnamefont{Kiselev}},
  \bibinfo{author}{\bibfnamefont{V.~G.} \bibnamefont{Chigrinov}},
  \bibnamefont{and} \bibinfo{author}{\bibfnamefont{D.~D.} \bibnamefont{Huang}},
  \bibinfo{journal}{Phys. Rev. E} \textbf{\bibinfo{volume}{72}},
  \bibinfo{pages}{061703} (\bibinfo{year}{2005}).

\bibitem[{\citenamefont{Chung et~al.}(2002)\citenamefont{Chung, Fukuda,
  Takanishi, Ishikawa, Matsuda, Takezoe, and Osipov}}]{Osipov:jap:2002}
\bibinfo{author}{\bibfnamefont{D.-H.} \bibnamefont{Chung}},
  \bibinfo{author}{\bibfnamefont{T.}~\bibnamefont{Fukuda}},
  \bibinfo{author}{\bibfnamefont{Y.}~\bibnamefont{Takanishi}},
  \bibinfo{author}{\bibfnamefont{K.}~\bibnamefont{Ishikawa}},
  \bibinfo{author}{\bibfnamefont{H.}~\bibnamefont{Matsuda}},
  \bibinfo{author}{\bibfnamefont{H.}~\bibnamefont{Takezoe}}, \bibnamefont{and}
  \bibinfo{author}{\bibfnamefont{M.~A.} \bibnamefont{Osipov}},
  \bibinfo{journal}{J. Appl. Phys.} \textbf{\bibinfo{volume}{92}},
  \bibinfo{pages}{1841} (\bibinfo{year}{2002}).

\bibitem[{\citenamefont{Huang and Rosenblatt}(2005)}]{Huang:apl:2005}
\bibinfo{author}{\bibfnamefont{Z.}~\bibnamefont{Huang}} \bibnamefont{and}
  \bibinfo{author}{\bibfnamefont{C.}~\bibnamefont{Rosenblatt}},
  \bibinfo{journal}{Appl. Phys. Lett.} \textbf{\bibinfo{volume}{86}},
  \bibinfo{pages}{011908} (\bibinfo{year}{2005}).

\bibitem[{\citenamefont{Chen et~al.}(1996)\citenamefont{Chen, Johnson, Bos,
  Wang, and West}}]{Chen:1996}
\bibinfo{author}{\bibfnamefont{J.}~\bibnamefont{Chen}},
  \bibinfo{author}{\bibfnamefont{D.~L.} \bibnamefont{Johnson}},
  \bibinfo{author}{\bibfnamefont{P.~J.} \bibnamefont{Bos}},
  \bibinfo{author}{\bibfnamefont{X.}~\bibnamefont{Wang}}, \bibnamefont{and}
  \bibinfo{author}{\bibfnamefont{J.~L.} \bibnamefont{West}},
  \bibinfo{journal}{Phys. Rev. E} \textbf{\bibinfo{volume}{54}},
  \bibinfo{pages}{1599} (\bibinfo{year}{1996}).

\bibitem[{\citenamefont{Kiselev}(2002)}]{Kis:jpcm:2002}
\bibinfo{author}{\bibfnamefont{A.~D.} \bibnamefont{Kiselev}},
  \bibinfo{journal}{J. Phys.: Condens. Matter} \textbf{\bibinfo{volume}{14}},
  \bibinfo{pages}{13417} (\bibinfo{year}{2002}).

\bibitem[{\citenamefont{Yaroshchuk et~al.}(2003)\citenamefont{Yaroshchuk,
  Kiselev, Zakrevskyy, Bidna, Kelly, Chien, and Lindau}}]{Kis:pre:2003}
\bibinfo{author}{\bibfnamefont{O.~V.} \bibnamefont{Yaroshchuk}},
  \bibinfo{author}{\bibfnamefont{A.~D.} \bibnamefont{Kiselev}},
  \bibinfo{author}{\bibfnamefont{Y.}~\bibnamefont{Zakrevskyy}},
  \bibinfo{author}{\bibfnamefont{T.}~\bibnamefont{Bidna}},
  \bibinfo{author}{\bibfnamefont{J.}~\bibnamefont{Kelly}},
  \bibinfo{author}{\bibfnamefont{L.-C.} \bibnamefont{Chien}}, \bibnamefont{and}
  \bibinfo{author}{\bibfnamefont{J.}~\bibnamefont{Lindau}},
  \bibinfo{journal}{Phys. Rev. E} \textbf{\bibinfo{volume}{68}},
  \bibinfo{pages}{011803} (\bibinfo{year}{2003}).




\end{thebibliography}

\end{document}